\documentclass[a4paper,11pt]{article}
\pdfoutput=1 
\usepackage{jcappub}

\usepackage{amsmath}
 
\usepackage{bm}
\usepackage{graphicx}
\usepackage{enumitem}
\usepackage{geometry}
\usepackage{xspace}
\usepackage{pdflscape}
\usepackage{placeins}
\usepackage{epstopdf}
\usepackage{comment}
\usepackage{subcaption}

\mathchardef\mhyphen="2D

\makeatletter
\gdef\@fpheader{}
\g@addto@macro\bfseries{\boldmath}
\makeatother

\allowdisplaybreaks

\newcommand{\ie}{{i.e.~}}

\newcommand{\eg}{e.g.~}

\let\oldsqrt\sqrt
\def\sqrt{\mathpalette\DHLhksqrt}
\def\DHLhksqrt#1#2{%
\setbox0=\hbox{$#1\oldsqrt{#2\,}$}\dimen0=\ht0
\advance\dimen0-0.2\ht0
\setbox2=\hbox{\vrule height\ht0 depth -\dimen0}%
{\box0\lower0.4pt\box2}}

\newcommand{\order}[1]{\mathcal{O}\!\left(#1\right)}

\newcommand{\dd}{\mathrm{d}}
\newcommand{\ee}{e}

\newcommand{\sss}[1]{{\scriptscriptstyle{#1}}}

\newcommand{\uPl}{\mathrm{Pl}}

\newcommand{\umax}{\mathrm{max}}
\newcommand{\uend}{\mathrm{end}}

\newcommand{\usssPl}{\sss{\uPl}}

\newcommand{\calP}{\mathcal{P}}

\newcommand{\GeV}{\mathrm{GeV}}

\newcommand{\cs}{c_{_\mathrm{S}}}

\newcommand{\Mp}{M_\usssPl}

\newcommand{\efolds}{$e$-folds}

\newcommand{\beq}{\begin{equation}}
\newcommand{\eeq}{\end{equation}}
\newcommand{\bea}{\begin{equation}\begin{aligned}}
\newcommand{\eea}{\end{aligned}\end{equation}}

\newlength{\wsingfig}
\newlength{\wdblefig}
\newlength{\wquadfig}
\newlength{\wtriplefig}
\newcommand{\Eq}[1]{Eq.~(\ref{#1})}
\newcommand{\Eqs}[1]{Eqs.~(\ref{#1})}
\newcommand{\Fig}[1]{Fig.~{\ref{#1}}}

\newcommand{\Refa}[1]{Ref.~{\cite{#1}}}
\newcommand{\Refs}[1]{Refs.~{\cite{#1}}}
\newcommand{\Sec}[1]{Sec.~\ref{#1}}

\newcommand{\App}[1]{Appendix~\ref{#1}}

\def\setI{\mathbb{I}}

\subheader{}

\title{Real-space entanglement in the Cosmic Microwave Background}

\author[a]{J\'er\^ome Martin,}
\author[b,a]{Vincent Vennin}

\affiliation[a]{Institut d'Astrophysique de Paris, UMR 7095-CNRS,
  Universit\'e Pierre et Marie Curie, 98 bis boulevard Arago, 75014 Paris, France}
\affiliation[b]{Laboratoire Astroparticule et Cosmologie,
  CNRS Universit\'e de Paris, 75013 Paris, France}

\emailAdd{jmartin@iap.fr}
\emailAdd{vincent.vennin@apc.univ-paris7.fr}

\date{today}

\begin{document}

\sloppy

\abstract{We compute the entanglement entropy, mutual information and quantum discord of the Cosmic Microwave Background (CMB) fluctuations in real space. To that end, we first show that measurements of these fluctuations at two distinct spatial locations can be described by a bipartite, continuous Gaussian system. This leads to explicit formulas for the mutual information and the quantum discord in terms of the Fourier-space power spectra of the curvature perturbation. We then find that quantum entanglement, that builds up in Fourier space between opposite wave momenta as an effect of quantum squeezing, is transferred to real space. In particular, both the mutual information and quantum discord, which decay as the fourth power of the distance between the two measurements in flat space time, asymptotes a constant in cosmological backgrounds. At the scales probed in the CMB however, they are highly suppressed, while they can reach order-one values at much smaller scales, where primordial black holes could have formed.
}


\maketitle

\section{Introduction}
\label{sec:intro}

The developments of quantum information theory over the past few
decades have given birth to various tools to characterise the presence
of genuine quantum correlations in multipartite systems, both for
discrete and continuous setups (see for instance
\Refa{2014arXiv1401.4679A} and references therein). Such tools are
useful to envision new applications of quantum systems in experimental
or industrial contexts, but also to shed some light in situations
where the very quantum nature of the physical process of interest is
under question.

This is notably the case in cosmology, where the structures observed
in the universe are understood as coming from the gravitational
amplification of quantum vacuum
fluctuations~\cite{Starobinsky:1979ty,Mukhanov:1981xt, Hawking:1982cz,
  Starobinsky:1982ee, Guth:1982ec, Bardeen:1983qw} during an era of
early accelerated expansion called inflation~\cite{Starobinsky:1980te,
  Sato:1980yn, Guth:1980zm, Linde:1981mu, Albrecht:1982wi,
  Linde:1983gd}. Although this mechanism leads to predictions that are
in excellent agreement with observations, such as the temperature and
polarisation anisotropies of the Cosmic Microwave Background
(CMB)~\cite{Akrami:2018odb}, it is conceptually not trivial since it
relies on quantising fluctuations of the metric together with the
matter content of the universe (and thus assumes the so-called linear
quantum-gravity approach), at energy scales that can be as high as
$10^{16}\, \GeV$, where quantum mechanics has never been tested so
far. It also leads to an exacerbated ``quantum-measurement''
problem~\cite{Sudarsky:2009za} related to how the quantum state of
cosmological structures acquired a collapsed configuration in the
early universe, which has far-fetching implications for quantum
mechanics itself (see \eg \Refa{Bassi:2003gd}). For these reasons, it
is important to better probe the possible quantum nature of
cosmological perturbations.

Since cosmological perturbations are described by a quantum field
evolving on a curved background, this implies to first extend the
relevant quantum-information tools to the realm of quantum
fields~\cite{Casini:2008wt, 2009PhRvA..80e2304D, Shiba:2012np, Lim:2014uea, Martin:2015qta,
  Kanno:2016gas, Bianchi:2017kgb, Espinosa-Portales:2019peb,
  Espinosa-Portales:2020pjp}. At leading order in perturbation theory,
cosmological perturbations are described by a free field (\ie without
non-linear interactions), and since it evolves on a homogeneous
background, its dynamics factorises in Fourier space. More precisely,
the field can be seen as an infinite set of uncoupled and independent
bipartite systems $\{\vec{k},-\vec{k}\}$, within which entangled pairs
of particles with opposite wave momenta are created due to the
presence of a strong external gravitational field. This leads to
large entanglement entropy between the sectors $\vec{k}$ and
$-\vec{k}$, which can be better characterised by computing the mutual information and quantum
discord~\cite{Henderson:2001,Zurek:2001} between $\vec{k}$ and
$-\vec{k}$. This has been done in \Refa{Martin:2015qta}, where it was
shown that the quantum discord indeed grows logarithmically with the
number of created pairs of quanta, \ie linearly with the phase-space
squeezing amplitude of the state, \ie again linearly with the number
of \efolds~spent outside the Hubble radius by the mode $k$ under
consideration. This allows one to reach very large values for the
quantum discord at the end of inflation, for the scales probed in the
CMB.

However, the fact that a large entanglement entropy, or quantum
discord, is found in Fourier space~\cite{Martin:2016tbd, Martin:2017zxs, Ando:2020kdz}, does not directly tell us how to
reveal its presence experimentally. Indeed, in practice, measurements
are performed in real space. Facing this situation, it is therefore interesting to study whether the presence of discord in Fourier space implies the presence of discord in real space and, if so, how efficiently it is transferred from one space to the other.

The problem of having characterised quantum correlations in Fourier space only, and the importance of investigating how it is related to quantum discord in real space, can also be illustrated with the example of Bell's inequality. The derivation of Bell's 
inequality (in its CHSH formulation) usually assumes 
that two observers, Alice and Bob, that are spatially separated, measure 
two dichotomic variables $A(\vec{a};\lambda)=\pm 1$ and $B(\vec{b};\lambda)=\pm 1$, respectively. In these expressions, $\vec{a}$ and $\vec{b}$ represent the settings of the detectors (typically the direction of a polariser).  The quantity $\lambda$ corresponds to ``hidden variables'', that is to say a set of variables that cannot be directly probed but that could influence the results read by Alice and/or Bob. In a classical theory, the mean value of $A(\vec{a};\lambda)$ is given by
\begin{align}
    \mathbb{E}_\lambda \left[A(\vec{a},\lambda)\right]=\int \dd A
    \, A(\vec{a},\lambda)\, p(A;\vec{a},\lambda) , 
\end{align}
where $p(A;\vec{a},\lambda)$ is the probability density function associated to the variable $A$. In a similar fashion, the two-point function can be expressed as
\begin{align}
    \mathbb{E}_\lambda \left[A\left(\vec{a},\lambda\right)
    B\left(\vec{b},\lambda\right)\right]
    =\int \dd A\, \dd B\, 
    \, A\left(\vec{a},\lambda\right)\, B\left(\vec{b},\lambda\right)
    p\left(A,B;\vec{a},\vec{b},\lambda\right) , 
\end{align}
where $p(A,B;\vec{a},\vec{b},\lambda)$ is the joint probability distribution associated to the measurement of $A$ and $B$. Then, if the theory is formulated in real space and, moreover, if it is local ``\`a la Bell'', what happens at Alice's location cannot influence what happens at Bob's location and vice-versa. Mathematically, this means that the joint distribution factorises, namely  $p(A,B;\vec{a},\vec{b},\lambda)=
p(A;\vec{a},\lambda) p(B;\vec{b},\lambda)$ which implies that 
\begin{align}
    \mathbb{E}_\lambda \left[A\left(\vec{a},\lambda\right)
    B\left(\vec{b},\lambda\right)\right]
= \mathbb{E}_\lambda \left[A(\vec{a},\lambda)\right]   
   \mathbb{E}_\lambda \left[B(\vec{b},\lambda)\right] .
    \end{align}
From this ``locality'' property, straightforward manipulations lead to the following inequality
\begin{align}
\label{eq:Bell}
\biggl \vert \mathbb{E}_\lambda \left[A\left(\vec{a},\lambda\right)
    B\left(\vec{b},\lambda\right)\right]
    &+\mathbb{E}_\lambda \left[A\left(\vec{a},\lambda\right)
    B\left(\vec{b}',\lambda\right)\right]+
    \mathbb{E}_\lambda \left[A\left(\vec{a}',\lambda\right)
    B\left(\vec{b},\lambda\right)\right]
    \nonumber \\ &
    -
    \mathbb{E}_\lambda \left[A\left(\vec{a}',\lambda\right)
    B\left(\vec{b}',\lambda\right)\right] \biggr \vert <2,
\end{align}
that is to say Bell's inequality. This inequality is important because, in quantum mechanics, the above quantity can be larger than $2$ (but is less than $2\sqrt{2}$, the so-called Cirelson's bound). Therefore, if observations indicate that the result is larger than $2$ (and, as is well-known, this does happen), then one has learned something deep about the natural world, namely that it can be ``non local''. At this point, it is worth stressing that this line of arguments relies on two, equally important, properties, namely (i) the fact that quantum mechanics can lead to physical situations where \Eq{eq:Bell} is violated, and (ii) the fact that, classically, this is not the case.
This is also why being able to factorise the joint distribution is crucial: without this ability, the quantity in \Eq{eq:Bell} could a priori take any value, so the second property mentioned above would not be verified and the fact that, in quantum mechanics, one can violate \Eq{eq:Bell}, would therefore be irrelevant. 

Let us now examine how the same problem is formulated in cosmology~\cite{Campo:2005sv, Maldacena:2015bha, Kanno:2017dci, Martin:2016tbd, Choudhury:2016cso, Martin:2017zxs, Ando:2020kdz}, when the analysis is carried out in Fourier space. What play the roles of Alice and Bob are the modes $+\vec{k}$ and $-\vec{k}$, and what 
play the roles of $A(\vec{a},\lambda)$ and $B(\vec{b},\lambda)$ are the variables $S_{\vec{k}}(\vec{m},\lambda)$ and $S_{-\vec{k}}(\vec{n},\lambda)$. These variables can be chosen according to different specifications but, for instance, in \Refs{Martin:2016tbd,Martin:2017zxs}, they are taken to be the so-called pseudo-spin operators which are dichotomic variables. At this stage, it is therefore possible to mimic in Fourier space the standard Alice-and-Bob approach described above. However, in order to construct a quantity which, classically, only takes values less than two, one needs 
to postulate the factorisation $p(S_{\vec{k}},S_{-\vec{k}}, \vec{m},\vec{n},\lambda )=p(S_{\vec{k}},\vec{m},\lambda) p(S_{-\vec{k}},\vec{n},\lambda)$. As mentioned above, for the case of Alice and Bob, this is based on the fact that these two observers are spatially separated and locality can be used to justify the factorisation. However, this reasoning does not hold for the modes $\vec{k}$ and $-\vec{k}$ since there is no concept of locality in Fourier space. As a consequence, even though, in principle, we can construct a quantity that would necessarily be less than 2 classically by postulating factorisation of the joint distribution in Fourier space, its physical justification is questionable.

In order to circumvent this problem, it is thus necessary to go from Fourier space to real space and to formulate the question of the quantum origin of the primordial perturbations in that space, where the notion of locality is meaningful. We have seen that, in Fourier space, there is a large entanglement entropy and a large amount of quantum discord. A first step in the program sketched before, which constitutes the main question investigated in the present article, is therefore to study whether the discord present in Fourier space is transferred to real space. The presence of discord in real space represents indeed a minimal requirement (a necessary condition) for our ability to reveal the quantum properties of the CMB fluctuations. This is why, in this work, we perform a generic calculation of the
mutual information and the quantum discord between measurements of a
free quantum field at distinct \emph{spatial} locations, before
applying our framework to the case of cosmological perturbations. Notice that the real-space mutual information has also been recently studied in \Refa{Espinosa-Portales:2020pjp} in a cosmological setting.

Let us finally mention that, in the case of standard Bell inequalities, an experimental violation requires to measure two non-commuting operators. In general, in a cosmological context, this implies to access the decaying mode, see \eg~\Refs{dePutter:2019xxv,Martin:2017zxs}, which may be possible in some specific models, see for instance \Refa{Maldacena:2015bha}, but is otherwise very challenging. However, the above discussion was based on the Bell inequality for illustrative purpose only. There are other quantum tests that may not be plagued with the same decaying-mode obstruction (for instance Leggett-Garg inequalities, as studied in \Refa{Martin:2016nrr}, or bipartite temporal Bell inequalities, see \Refa{Ando:2020kdz}). The relevance of going from Fourier to real space applies broadly and motivates our work beyond the mere application to Bell inequalities. This is also the reason why we compute the quantum discord, which is a generic tool to assess the presence of quantum correlations (independently of a concrete experimental test to reveal them). It is clear that the Bell inequality is not the only mean to reveal the presence of quantum properties and one can easily imagine that a non-vanishing discord could be tested by other means. In fact, the present study might precisely point towards other ideas to probe the quantum nature of cosmological perturbations. Let us moreover mention that the approach presented here may also be relevant for other systems (possibly in the lab), where the above-mentioned limitation does not apply.

The paper is organised as follows. In \Sec{sec:TwoPointFunction}, we
explain how two-point measurements of a free quantum field, at spatial
locations $\vec{x}_1$ and $\vec{x}_2$, can be described in terms of a
Gaussian bipartite system. Such systems are fully characterised by
their covariance matrix, which we relate to the Fourier-space power
spectra of the field. In \Sec{sec:MutualInformation:Discord}, we
explain how the mutual information and the quantum discord can be
computed for Gaussian bipartite systems from the entries of the
covariance matrix. This allows us to establish generic formulas that
can be used for any free quantum field. In \Sec{sec:Cosmo}, we apply
our framework to the case of cosmological perturbations, during the
early epoch of inflation as well as during the subsequent era where
the universe is dominated by a radiation fluid. We present numerical
results obtained by evaluating the formulas derived in the
previous two sections, but most of \Sec{sec:Cosmo} is devoted to the
derivation of analytical approximations. Those approximations are then
summarised in \Fig{fig:summary} in \Sec{sec:conclusions}, where they
are further commented on and where we present our main
conclusions. Finally, the paper ends with \App{app:Integrals} to which
technical details of the approximation performed in \Sec{sec:Cosmo}
are deferred.
\section{Bipartite systems for two-point measurements of a quantum field}
\label{sec:TwoPointFunction}

\subsection{General considerations}
\label{subsec:generalcons}

Let $\phi(\vec{x})$ be a (classical random or quantum) Gaussian real
field, and $\pi(\vec{x})$ its conjugated momentum, arranged into the
vector $\bm{z}(\vec{x}) = (\phi(\vec{x}),
\pi(\vec{x}))^\mathrm{T}$. They can be expanded into Fourier moments
according to
\begin{align}
\label{eq:Fourier:transform}
z_i(\vec{x}) = \frac{1}{\left(2\pi\right)^{3/2}} \int \dd^3 \vec{k}
\, \ee^{-i \vec{k} \cdot \vec{x}} z_i(\vec{k}) \qquad \text{where}\qquad
i=1,2\, ,
\end{align} and where the reality condition
$\bm{z}(\vec{x})=\bm{z}^\dagger(\vec{x})$ imposes that
$\bm{z}(-\vec{k}) = \bm{z}^\dagger(\vec{k})$. The fields $\phi$
  and $\pi$ are canonically conjugated one to another, which implies
  that they satisfy
\begin{align}
\label{eq:commutators:RealSpace}
\left[\phi(\vec{x}_1),\pi(\vec{x}_2)\right] = i \delta(\vec{x}_1-\vec{x}_2)\, ,
\end{align}
where $[A,B]\equiv AB-BA$ denotes the quantum commutator in the case of quantum
fields (the case of classical fields can be treated similarly by
replacing quantum commutators by Poisson brackets). The above
commutation relation can be written in a matricial form for any pair of entries of the vector ${\bm z}$ as follows,
\begin{align}
\label{eq:commutators:RealSpace:MatricialForm}
  \left[z_i(\vec{x}_1),z_j(\vec{x}_2)\right] = i J_{ij}^{(1)}
  \delta\left(\vec{x}_1-\vec{x}_2\right)
\qquad \text{where}\qquad
\bm{J}^{(1)}=
\begin{pmatrix}
0 & 1 \\
-1 & 0
\end{pmatrix}
\, .
\end{align}
Moreover, making use of \Eq{eq:Fourier:transform}, one has
$\left[\phi(\vec{k}_1),\pi^\dagger(\vec{k}_2)\right] = i
\delta(\vec{k}_1-\vec{k}_2)$, and this leads to the same commutation
relations in Fourier space,
\begin{align}
  \left[z_i(\vec{k}_1),z_j^\dagger(\vec{k}_2)\right] = i J_{ij}^{(1)}
\delta(\vec{k}_1-\vec{k}_2)\, ,
\end{align}
\ie the Fourier transform~\eqref{eq:Fourier:transform} is a canonical
transformation.

The fields $\phi(\vec{x})$ and $\pi(\vec{x})$ being Gaussian, their
statistical properties are entirely determined by their two-point
correlation functions
\begin{align}
\label{PowerSpectrum:interm1}
\left\langle \left\lbrace z_i(\vec{x}_1),  z_j (\vec{x}_2) \right\rbrace
\right\rangle = \frac{1}{\left(2\pi\right)^3} \int\dd^3 \vec{k_1}
\int \dd^3 \vec{k_2} \, \ee^{i \vec{k}_1\cdot \vec{x}_1
  - i \vec{k}_2\cdot \vec{x}_2} \left\langle
\left\lbrace z_i^\dagger(\vec{k}_1),z_j(\vec{k}_2)\right\rbrace\right\rangle\, ,
\end{align}
where $\langle \cdot \rangle$ denotes quantum expectation value (or
statistical average in the case of classical random fields), and
$\{A,B\}\equiv (AB+BA)/2$ stands for half of the anticommutator. Hereafter we assume that the fields $\phi$ and
  $\pi$ are placed in a configuration that is statistically
  homogeneous and isotropic, \ie it is invariant under spatial
  translations and rotations. This is the case if the Hamiltonian that
  drives their dynamics enjoys the same symmetries, as in the context
  of Friedmann-Lema\^itre-Robertson-Walker (\ie homogeneous and
  isotropic) cosmologies for instance. Invariance under translations
  implies that $\langle \{ z_i(\vec{x}_1), z_j (\vec{x}_2)
  \}\rangle=\langle\{ z_i(\vec{x}_1+\vec{D}) , z_j
  (\vec{x}_2+\vec{D})\} \rangle$ for any displacement vector
  $\vec{D}$. From \Eq{PowerSpectrum:interm1}, this condition leads to
  $\langle \{z_i^\dagger(\vec{k}_1),z_j(\vec{k}_2)\}\rangle =\langle
  \{z_i^\dagger(\vec{k}_1),z_j(\vec{k}_2)\}\rangle \ee^{i
    (\vec{k}_1-\vec{k}_2)\cdot \vec{D}}$ for all $\vec{D}$, which can
  only be satisfied if $\langle \{
  z_i^\dagger(\vec{k}_1),z_j(\vec{k}_2)\}\rangle$ involves a Dirac
  function $\delta(\vec{k_1}-\vec{k}_2)$, \ie if
\begin{align}
\label{PowerSpectrum:interm2}
\left\langle \left\lbrace z_i^\dagger(\vec{k}_1),
z_j(\vec{k}_2)\right\rbrace \right\rangle
= \frac{2\pi^2}{k_1^3}\calP_{ij} (\vec{k}_1)
\delta(\vec{k_1}-\vec{k}_2)\, .
\end{align}
This expression defines the reduced power spectrum
$\calP_{ij}(\vec{k})$, where the prefactor $2\pi^2/k^3$ has been introduced for later convenience. Invariance under rotations then
implies that $\langle \{ z_i(\vec{x}_1) , z_j
(\vec{x}_2)\}\rangle=\langle \{ z_i[R(\vec{x}_1)] , z_j
[R(\vec{x}_2)]\}\rangle$ for any rotation operator $R$. As a
consequence, \Eq{PowerSpectrum:interm1} leads to $\langle
\{z_i^\dagger(\vec{k}_1),z_j(\vec{k}_2)\}\rangle = \langle \{
z_i^\dagger[R^{-1}(\vec{k}_1)], z_j[R^{-1}(\vec{k}_2)]
\}\rangle$. Using \Eq{PowerSpectrum:interm2}, this translates into
$\calP_{ij} (\vec{k}) = \calP_{ij} [R^{-1}(\vec{k})] $ for all
rotation operators $R$, hence $\calP_{ij}$ only depends on the norm
$k=\vert \vec{k}\vert$ of the vector $k$.

This leads to the reduced power-spectrum functions $\calP_{ij}(k)$, in
terms of which \Eq{PowerSpectrum:interm1} yields
\begin{align}
\label{eq:PowerSpectrum:RealSpace}
\left\langle \left\{z_i(\vec{x}_1), z_j (\vec{x}_2)\right\}\right\rangle
= \int_0^\infty \calP_{ij}(k) \,
\mathrm{sinc}\left(k \left\vert \vec{x}_1
- \vec{x}_2\right\vert\right)\dd\ln k\, ,
\end{align}
where $\mathrm{sinc}\, x=\sin x/x$ is the cardinal sine function, and
which determine all observables in general. 
\subsection{Coarse graining}
\label{subsec:Coarse:Graining}
In practice, any measurement device has a finite spatial resolution
that we denote $R$. This means that experiments probing the fields
$\phi(\vec{x})$ and $\pi(\vec{x})$ are only sensitive to their value
locally averaged over a patch of size $R$, see also
Fig.~\ref{fig:sketchscale}. This leads us to introduce the
coarse-grained fields
\begin{align}
\label{eq:def:CoarseGrain}
z_{R,i} (\vec{x}) \equiv \left(\frac{a}{R}\right)^3
\int\dd^3\vec{y} \, z_i (\vec{y})\, 
W\left(\frac{a\left\vert \vec{y} - \vec{x}\right\vert}{R} \right) .
\end{align}
One notes that a (time-dependent) scale factor $a(t)$ has been
introduced in this expression, in order to make our formalism directly
applicable to the cosmological setting in \Sec{sec:Cosmo}. In
cosmology indeed, it is convenient to work with the so-called
``comoving'' spatial coordinate $\vec{x}$, related to the ``physical''
coordinate $\vec{x}_{\mathrm{p}}= a(t) \, \vec{x}$. In the above
expression, $\vec{x}$ and $\vec{y}$ are therefore comoving while $R$
denotes a physical distance. If one is not interested in cosmological
applications, one can simply set $a=1$ in \Eq{eq:def:CoarseGrain} and
in all following formulas, since the formalism presented in this section is generic (and will be specified to the cosmological setting in \Sec{sec:Cosmo} only).  In \Eq{eq:def:CoarseGrain}, $W$ is a
window function that depends only on the distance away from $\vec{x}$
in order to preserve isotropy, and which satisfies $W(x)\simeq 1$ if $x\ll
1$ and $W(x)\simeq 0 $ if $x\gg 1$. It is normalised such that
\begin{align}
\label{eq:Normalisation:Cond:W}
\int_0^\infty x^2 W(x)\, \dd x=\frac{1}{4\pi}\, ,
\end{align}
\ie such that after coarse graining, a uniform field remains a uniform
field of the same value.

Upon Fourier transforming \Eq{eq:def:CoarseGrain}, one obtains
\begin{align}
\label{eq:def:tilde:W}
z_{R,i} (\vec{k}) = 
z_i(\vec{k})
4\pi
\left(\frac{a}{kR}\right)^3 \int_0^\infty W\left(\frac{a u}{kR} \right)
u \sin u \, \dd u
\equiv z_i(\vec{k})\widetilde{W}\left(\frac{k R}{a}\right)\, ,
\end{align}
which defines $\widetilde{W}$, that shares similar properties with
$W$. Indeed, when $a/(kR) \gg 1$, the values of $u$ such that
$W\left[a u/(k R) \right]$ is not close to zero are much smaller than
one, so one can replace $\sin u\simeq u$ in the integral over $u$, and
using the normalisation condition~\eqref{eq:Normalisation:Cond:W}, one
obtains $\widetilde{W}(k R/a) \simeq 1$ in that limit. In the opposite
limit where $a/(k R) \ll 1$, since $W\left[a u/(k R) \right]\simeq 1$
until $u\sim k R/a$, the integral over $u$ in \Eq{eq:def:tilde:W} is at most
of order $k R/a$, hence $\widetilde{W}\left(k R/a\right)\lesssim
[a/(kR)]^2\ll 1 $.

\begin{figure}[t]
\begin{center}
  \includegraphics[width=0.49\textwidth]{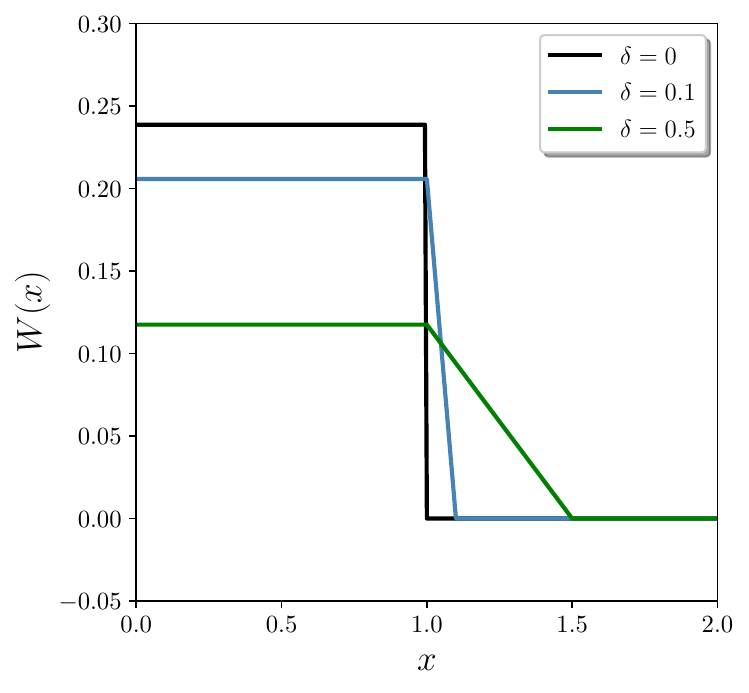}
  \includegraphics[width=0.49\textwidth]{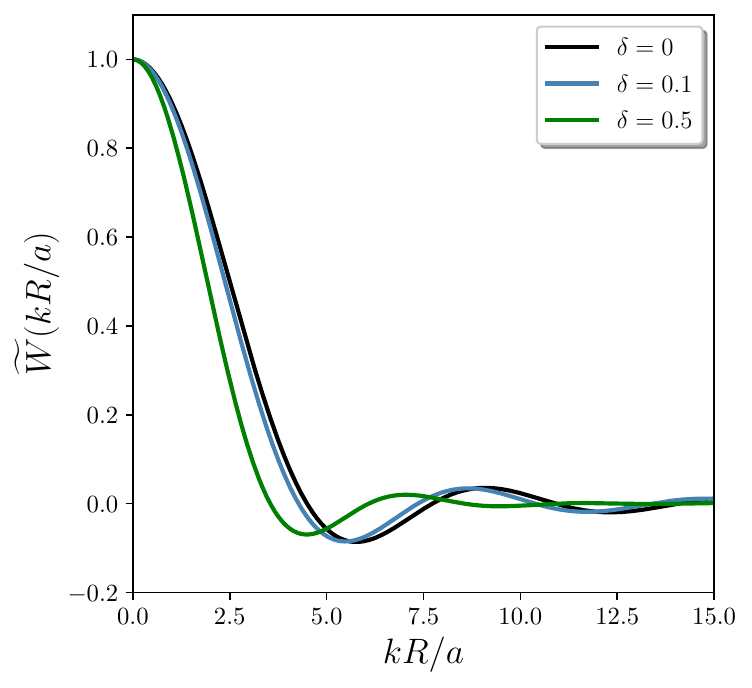}
  \caption{Left panel: window
    function~\eqref{eq:WindowFunction:Improved} in real space, for $\delta =0$ (black line), $\delta =0.1$
    (blue line) and $\delta =0.5$ (green line). The case $\delta=0$
    corresponds to \Eq{eq:W:Heaviside}. Right panel: window
    function in Fourier space, see
    \Eq{eq:Fourier:transform:Heaviside:Improved}, for the same
    values of $\delta$. The case $\delta=0$ is given by
    \Eq{eq:Fourier:transform:Heaviside}.}
\label{fig:windowrealfourier}
\end{center}
\end{figure}

The details of $\widetilde{W}$ between these two limits depend on
those of $W$. For instance, if $W$ is a Heaviside step function (see
the black line in the left panel of
Fig.~\ref{fig:windowrealfourier}),
\begin{align}
\label{eq:W:Heaviside}
W(x) = \frac{3}{4\pi} \theta(1-x),
\end{align}
where $\theta(x)=1$ if $x>0$ and $0$ otherwise, and where the
pre-factor is set in such a way that the normalisation
condition~\eqref{eq:Normalisation:Cond:W} is satisfied,
\Eq{eq:def:tilde:W} gives rise to
\begin{align}
\label{eq:Fourier:transform:Heaviside}
\widetilde{W}\left(\frac{k R}{a}\right) = 3 \left(\frac{a}{k R}\right)^3
\left[ \sin \left(\frac{k R}{ a}\right)-\frac{k R}{ a}
  \cos\left(\frac{k R}{a }\right)\right] .
\end{align}
This verifies the two limits given in the main text and is
represented by the black line in the right panel of
\Fig{fig:windowrealfourier}. However, the
sharpness of the Heaviside profile~\eqref{eq:W:Heaviside} in real
space (namely the fact that $W$ is not a continuous function) leads to
mild UV divergences in some of the intermediate quantities we compute below. This suggests
to use a smoother window function such as
\begin{align}
\label{eq:WindowFunction:Improved}
W(x)=\frac{3}{4\pi {\cal F}(\delta)}
\begin{cases}
1 \qquad \text{for} \qquad x\le 1\, ,\\
\displaystyle
-\frac{1}{\delta}(x-1)+1 \qquad \text{for}\qquad 1<x\le 1+\delta\, , \\
0 \qquad \text{for}\qquad x>1+\delta \, ,
\end{cases}
\end{align}
where
\begin{align}
  {\cal F}(\delta)=\frac14(\delta+2)(\delta ^2+2\delta +2)
\end{align}
is set such that the normalisation
condition~\eqref{eq:Normalisation:Cond:W} is satisfied. This
generalises \Eq{eq:W:Heaviside}, which is recovered when $\delta=0$,
by adding a linear tail between $x=1$ and $x=1+\delta$ in order to
make $W$ continuous. This window function is represented in the left panel of
\Fig{fig:windowrealfourier}. From \Eq{eq:def:tilde:W}, one finds
\begin{align}
\label{eq:Fourier:transform:Heaviside:Improved}
  \widetilde{W}\left(\frac{kR}{a}\right)=&\frac{3}{{\cal F}(\delta)}\left(\frac{kR}{a}\right)^{-3}
  \biggl\{\frac{1}{\delta}\sin \left(\frac{kR}{a}\right)
  -\left(1+\frac{1}{\delta}\right)\sin \left[(1+\delta )\frac{kR}{a}\right]
  \nonumber \\ &
  +\frac{2}{\delta }\frac{a}{kR}\cos \left(\frac{kR}{a}\right)
  -\frac{2}{\delta}\frac{a}{kR}
  \cos \left[(1+\delta)\frac{kR}{a}\right]\biggr\}\, ,
\end{align}
which is represented in the right panel of
  \Fig{fig:windowrealfourier} for different values of $\delta$. One can check that this formula
  for the window function in Fourier space reduces to
\Eq{eq:Fourier:transform:Heaviside} in the limit $\delta\to 0$. One
can also see that, in the limit $kR/a\ll 1$, both
\Eqs{eq:Fourier:transform:Heaviside}
and~\eqref{eq:Fourier:transform:Heaviside:Improved} are such that
$\widetilde{W}\simeq 1$. However, in the regime $kR/a\gg 1$,
\Eq{eq:Fourier:transform:Heaviside} leads to $\widetilde{W}\simeq
a^2/(kR)^2$ while \Eq{eq:Fourier:transform:Heaviside:Improved} leads
to $\widetilde{W}\simeq a^3/(kR)^3$, which ensures UV convergence
for all the quantities of interest below.

Let us also note that other smooth functions could have been
employed for $W$, for instance a Gaussian profile as often done, but
as we are now going to see, the window function needs to
have a compact support in order for a bipartite system to be defined with canonical commutation relations, and this makes the above choice
natural. Other smooth, yet compact, window functions could obviously
be considered (and tailored to better model a given experiment's
measuring device), but this would only lead to small and irrelevant
modifications of the results presented in the following.

The next step consists in checking that the commutation relations~\eqref{eq:commutators:RealSpace:MatricialForm} are
still satisfied after coarse graining, that is to say, one should
check that the coarse-graining procedure is a canonical transformation
of the phase space. It is clear that one still has
$[\phi_R(\vec{x}_1),\phi_R(\vec{x}_2)]=[\pi_R(\vec{x}_1),\pi_R(\vec{x}_2)]=0$. Making
use of \Eqs{eq:def:CoarseGrain} and~\eqref{eq:commutators:RealSpace},
one finds
 \begin{align}
 \label{eq:commutator:Interm}
 \left[ \phi_R(\vec{x}_1), \pi_R(\vec{x}_2)\right]
 = i \left(\frac{a}{R}\right)^6 \int \dd^3 \vec{y}~
 W\left(\frac{a}{R}\left\vert \vec{y}-\vec{x}_1\right\vert\right)
 W\left(\frac{a}{R}\left\vert \vec{y}-\vec{x}_2\right\vert\right)\, .
 \end{align}
 Note that if the
 support of the window function is not bounded in real space, then the
 above integral is necessarily strictly positive and it is clear that
 one cannot get $ \left[ \phi_R(\vec{x}_1), \pi_R(\vec{x}_2)\right]
 =0$. As already mentioned, this is the reason why a compact window
 function was previously introduced, and hereafter, we make use of
 \Eq{eq:WindowFunction:Improved} for explicitness. The
 commutator~\eqref{eq:commutator:Interm} then vanishes if the two
 patches do not overlap, \ie if the two spatial points are
 sufficiently distant away,
\begin{align} 
\label{eq:lowerbound:d}
d\equiv a \vert \vec{x}_1 - \vec{x}_2 \vert>2 R (1+\delta)\, ,
\end{align}
where $d=a \vert \vec{x}_1 - \vec{x}_2 \vert$ denotes the (physical)
distance between $\vec{x}_1$ and $\vec{x}_2$, see also
\Fig{fig:sketchscale}. In the coincident limit,
$\vec{x}_2=\vec{x}_1$, \Eq{eq:commutator:Interm} gives rise to $ [
  \phi_R(\vec{x}), \pi_R(\vec{x})] = 4i\pi \left(a/R\right)^3 \int \dd
u W^2(u) u^2$. Together with \Eq{eq:WindowFunction:Improved}, this
leads to
\begin{align}
\label{eq:commutaror:cg}
  \left[ \phi_R(\vec{x}_i), \pi_R(\vec{x}_j)\right]
  =& i \frac{3}{4\pi} \left(\frac{a}{R}\right)^3 G(\delta)\delta_{ij},
\end{align}
where $i,j=1,2$ and $\vec{x}_1$ and $\vec{x}_2$ satisfy \Eq{eq:lowerbound:d}, and where
\begin{align}
  G(\delta) =\frac{8 \left(\delta ^3+5 \delta ^2+10 \delta +10\right)}
  {5 (\delta +2)^2 \left(\delta ^2+2 \delta +2\right)^2}\, .
\end{align}
The prefactor in this expression has been arranged such that, when
$\delta=0$, $G(\delta)=1$. Since the
commutator~(\ref{eq:commutaror:cg}) differs from \Eq{eq:commutators:RealSpace:MatricialForm}, the fields need to be rescaled, and for this reason we introduce
\begin{align}
  \tilde{\bm z}_R={\bm \Lambda }^{(1)}{\bm z}_R,
\end{align}
with
\begin{align}
  {\bm \Lambda}^{(1)}\equiv \left(\frac{R}{a}\right)^{3/2}
  \sqrt{\frac{4\pi}{3 G(\delta)}}
  \begin{pmatrix}
    \lambda & 0 \\
    0 & \lambda^{-1}
  \end{pmatrix}.
  \end{align}
One can check that $\tilde{\bm z}_R$ is indeed canonically normalised, by explicitly calculating
\begin{align}
  \left[\tilde{z}_{R,i}(\vec{x}),\tilde{z}_{R,j}(\vec{x})\right]
  =&\Lambda_{ik}^{(1)}\left[z_{R,k}(\vec{x}),z_{R,\ell}(\vec{x})\right]
  \left(\Lambda^{(1)}{}^{_\mathrm{T}}\right)_{\ell j}
  =iJ_{ij}^{(1)}.
\end{align}
Note that we have introduced a new parameter $\lambda$, which serves
two purposes. First, it may be set in such a way that the entries of
the vector $\tilde{\bm{z}}_R $ share the same dimension (and,
conveniently, are dimensionless), which simplifies some of the
following calculations. Second, changing $\lambda$ amounts to
performing a phase-space dilatation, which is a special case of
canonical transformations. Since some of the quantities we compute in the
following are local-symplectic invariant, checking their
non-dependence on $\lambda$ will be a valuable sanity check.

Finally, it is interesting to calculate the two-point correlation
  function of the coarse-grained fields. Plugging \Eq{eq:def:tilde:W}
  into (the coarse-grained version of)
  \Eq{eq:PowerSpectrum:RealSpace}, one has
\begin{align}
  \left\langle \left\{z_{i,R}(\vec{x}_1),  z_{j,R}(\vec{x}_2)\right\}
  \right\rangle = 
\int_0^\infty \widetilde{W}^2\left(\frac{R}{a}k\right)
\calP_{ij}(k) \, \mathrm{sinc}\left(k \left\vert \vec{x}_1
- \vec{x}_2\right\vert\right)\dd\ln k\, ,
\end{align}
where we recall that $\widetilde{W}$ is given in
\Eq{eq:Fourier:transform:Heaviside:Improved}. This equation should
  be compared to \Eq{eq:PowerSpectrum:RealSpace}, to which it reduces in the limit $R\to 0$. The only difference
  is the presence of the squared window function, which originates from the fact that we have considered
  coarse-grained quantities.
\subsection{Bipartite system}
\label{subsec:bipartite}
Our goal is now to characterise the presence of entanglement between the
configurations of the fields at two different locations $\vec{x}_1$ and
$\vec{x}_2$. We therefore view our setup as a bipartite system,
containing the values of the coarse-grained fields as those two
locations, and arranged into the vector
\begin{align}
  \bm{Z}_R(\vec{x}_1,\vec{x}_2)=
  \begin{pmatrix}
    {\bm z}_R(\vec{x}_1) \\
    {\bm z}_R(\vec{x}_2)
  \end{pmatrix}
  =
\left(
\begin{array}{c}
{\phi}_R(\vec{x}_1)\\
{\pi}_R(\vec{x}_1)\\
{\phi}_R(\vec{x}_2)\\
{\pi}_R(\vec{x}_2)
\end{array}
\right)\, .
\end{align}
The two first entries of the vector $\bm{Z}_R(\vec{x}_1,\vec{x}_2)$
contain the phase-space variables of the ``first'' system, \ie the one
observed at location $\vec{x}_1$, while the two last entries contain
the phase-space variables of the ``second'' system, \ie the one
observed at location $\vec{x}_2$. In some sense, the vector ${\bm
  Z}_R(\vec{x}_1,\vec{x}_2)$ is an ``enlarged'' version of ${\bm
  z}_R(\vec{x})$, which explains the notation with a capital
letter. Its component will be denoted with a Latin letter, \ie
$Z_{R,a}(\vec{x}_1,\vec{x}_2)$ with $a=1, \cdots 4$.

It is clear that the vector ${\bm Z}_R(\vec{x}_1,\vec{x}_2)$ suffers
from the same issue as ${\bm z}_R$, namely that it is not canonically
normalised. This problem can be solved following the
considerations presented in \Sec{subsec:Coarse:Graining}, \ie by defining $\tilde{\bm Z}_R(\vec{x}_1,\vec{x}_2)=(\tilde{\bm
  z}_R(\vec{x}_1),\tilde{\bm z}_R(\vec{x}_2))^{_\mathrm{T}}$, or
\begin{align}
\label{eq:Field:Redef}
\tilde{\bm{Z}}_R(\vec{x}_1,\vec{x}_2)
\equiv  {\bm \Lambda}^{(2)}
 \bm{Z}_R(\vec{x}_1,\vec{x}_2)
 \end{align}
 where
 \begin{align}
 {\bm \Lambda}^{(2)}=\left(\frac{R}{a}\right)^{3/2} \sqrt{\frac{4\pi}{3 G(\delta)}}
\left(
\begin{array}{cccc}
\lambda & 0 & 0 & 0\\
0 & \lambda^{-1} & 0& 0\\
0 & 0& \lambda& 0\\
0&0&0& \lambda^{-1}
\end{array}
\right)={\bm \Lambda}^{(1)}\oplus{\bm \Lambda}^{(1)}\, .
\end{align}
From \Eq{eq:commutaror:cg}, the entries of
$\tilde{\bm{Z}}_R$ satisfy the following canonical commutation relations
\begin{align}
  \left[\tilde{Z}_{R,a}(\vec{x}_1,\vec{x}_2),
    \tilde{Z}_{R,b}(\vec{x}_1,\vec{x}_2)\right] = i J_{ab}^{(2)},
    \end{align}
    the matrix ${\bm J}^{(2)}$ being defined by
\begin{align}
\bm{J}^{(2)}=
\left(
\begin{array}{cccc}
0 & 1 & 0 & 0 \\
-1 & 0 & 0 & 0\\
0 & 0 & 0  & 1\\
0 & 0 & -1 & 0
\end{array}
\right)\, 
={\bm J}^{(1)}\oplus {\bm J}^{(1)} ,
\end{align}
where we recall that $\vec{x}_1$ and $\vec{x}_2$ must satisfy \Eq{eq:lowerbound:d}.
We have thus parametrised our bipartite system with canonical
coordinates, which was the aim of this subsection. 

\subsection{Covariance matrix}
\label{subsec:cov}
As argued above, the fields being Gaussian, they are entirely
described by their two-point statistics. For this reason, let us
introduce the covariance matrix $\bm{\gamma}$, defined by
\begin{align}
\label{eq:Covariance:matrix:def}
\left\langle \tilde{Z}_{R,a}(\vec{x}_1,\vec{x}_2)
\tilde{Z}_{R,b}(\vec{x}_1,\vec{x}_2)\right\rangle
= \frac{1}{2}\gamma_{ab}+\frac{i}{2}J_{ab}^{(2)}\, ,
\end{align}
which also implies that $\gamma_{ab}=2 \langle\{
\tilde{Z}_{R,a}(\vec{x}_1,\vec{x}_2) ,
\tilde{Z}_{R,b}(\vec{x}_1,\vec{x}_2) \}\rangle$. This leads to
\begin{align}
\label{eq:CorrelationMatrix:PowerSpectra}
\bm{\gamma} = &\frac{8\pi }{3G(\delta)}\left(\frac{R}{a}\right)^3
\int \dd\ln k \, \widetilde{W}^2\left(\frac{R}{a}k\right)
\nonumber \\ & \times
\left(
\begin{array}{cccc}
  \lambda^2 \calP_{\phi\phi}(k) & \calP_{\phi\pi}(k) &
  \displaystyle \lambda^2 \calP_{\phi\phi}(k)\,
  {\mathrm{sinc}\left(\frac{kd}{a} \right)}&
  \displaystyle \calP_{\phi\pi}(k)\, \mathrm{sinc}
  \left(\frac{k d}{a}\right)\\ \\
  \displaystyle
  - & \displaystyle \lambda^{-2} \calP_{\pi\pi}(k)  & \displaystyle
  \calP_{\phi\phi}(k)\, \mathrm{sinc}\left(\frac{k d}{a}\right)& \displaystyle
  \lambda^{-2} \calP_{\pi\pi}(k) \, \mathrm{sinc}\left(\frac{k d}{a}\right)\\ \\
  - & - & \lambda^2 \calP_{\phi\phi}(k)  & \calP_{\phi\pi}(k)
  \\ \\
 -  & - & - &  \lambda^{-2} \calP_{\pi\pi}(k) 
\end{array}
\right)\, ,
\end{align}
where the entries that are not explicitly written are obtained from
the symmetry of the covariance matrix,
$\bm{\gamma}=\bm{\gamma}^{T}$. The invariance of the system under
exchanging $\vec{x}_1$ and $\vec{x}_2$ also leads to an additional
symmetry\footnote{In the case where the sizes of the regions centred around $\vec{x}_1$ and $\vec{x}_2$ are different, \ie when $R_1\neq R_2$, this symmetry is lost, but the same method can still be employed, see \Refa{Martin:2021xml}.} of the covariance matrix (namely under the permutation matrix
that swaps the first and third, and the second and fourth, entries of
the vector $\tilde{\bm{Z}}_R$), such that there are only 6 independent entries
in the matrix $\bm{\gamma}$, that we label with $\gamma_{11}$,
$\gamma_{12}$, $\gamma_{22}$, $\gamma_{13}$, $\gamma_{14}$ and
$\gamma_{24}$. For instance, the determinant of the covariance matrix
is given by
\begin{align}
\label{eq:det:gamma}
\det \bm{\gamma} &=\left[(\gamma_{12}-\gamma_{14})^2
  -(\gamma_{11}-\gamma_{13}) (\gamma_{22}-\gamma_{24})\right]
\left[(\gamma_{12}+\gamma_{14})^2-(\gamma_{11}+\gamma_{13})
  (\gamma_{22}+\gamma_{24})\right]\, .
\end{align}
\section{Mutual information and quantum discord}
\label{sec:MutualInformation:Discord}
In \Sec{sec:TwoPointFunction}, we have seen how a Gaussian scalar
field measured at two distinct spatial locations $\vec{x}_1$ and
$\vec{x}_2$, coarse-grained over a spatial distance $R$, can be
described by a four-dimensional Gaussian state entirely specified by
the density matrix $\bm{\gamma}$, which is related to the power
spectra of the field via \Eq{eq:CorrelationMatrix:PowerSpectra}. Let
us now characterise the correlations that exist between measurements
performed at $\vec{x}_1$ and $\vec{x}_2$. This is done by means of two
quantities that play an important role in quantum information theory:
the mutual information that measures the amount of correlations, and
the quantum discord that measures the amount of \emph{quantum}
correlations. Our goal is to relate those two quantities to the
entries of the correlation matrix, which we have computed previously.
\subsection{Mutual information}
\label{sec:MutualInformation}
Let us first consider the case where $\bm{z}_R(\vec{x})$ is a classical
random field.  We formally denote by $\{a_i\}$ and $\{ b_j\}$ the
possible configurations of the field at the location $\vec{x}_1$ and
$\vec{x}_2$ respectively. We also introduce the probability $p_i$ to
find the field at $\vec{x}_1$ in the configuration $a_i$ and similarly
for $p_j$. The uncertainty regarding the state of the field at
$\vec{x}_1$ can be characterised by the von-Neumann entropy
\begin{align} S_1 =
  -\sum_{i} p_i \log_2(p_i)\, .
\end{align}
Indeed, if all $a_i$ vanish but one (so the configuration of the field
at $\vec{x}_1$ is certain), one can check that $S_1=0$, and that, in
general, $S_1\geq 0$. A similar expression for $S_2$ can be
introduced, as well as for the joint system
\begin{align}
  S_{1,2} = -\sum_{i,j} p_{ij} \log_2(p_{ij}),
\end{align}
where $p_{ij}$ denotes the joint probability to find the field at
$\vec{x}_1$ in configuration $a_i$ and at $\vec{x}_2$ in configuration
$b_j$. A measure of the mutual information between the configurations
at the two spatial locations is given by
\begin{align}
\label{eq:calI:def}
\mathcal{I}(\vec{x}_1,\vec{x}_2) = S_1+S_2-S_{1,2}\, .
\end{align}
The fact that $\mathcal{I}$ measures the presence of spatial
correlations can be seen by noting that if the two configurations are
uncorrelated, then the mutual information vanishes. Indeed, if
$p_{ij} = p_i p_j$, then $\mathcal{I} = -\sum_i p_i
\log_2(p_i)-\sum_j p_j \log_2(p_j) + \sum_{i,j} p_i p_j [\log_2(p_i) +
  \log_2(p_j)]=0$, where we have used that $\sum_i p_i = \sum_j p_j =
1$.

Let us now translate these considerations into the quantum formalism,
where our goal is to construct an analogue of $\mathcal{I}$. The full
quantum system can be described by its density matrix ${\rho}_{1,2}$,
and information about the field configuration at location $\vec{x}_1$
is obtained by tracing over the degrees of freedom corresponding to
$\vec{x}_2$, namely
\begin{align}
{\rho}_1 = \mathrm{Tr}_2 \left({\rho}_{1,2}\right)\, ,
\end{align}
and similarly for ${\rho}_2$. The state represented by ${\rho}_1$ is
still Gaussian, and its covariance matrix $\bm{\gamma}_1$ is simply
obtained from $\bm{\gamma}$ by removing the lines and columns
corresponding to $\vec{x}_2$, \ie the third and fourth lines and
columns, so
\begin{align}
\label{eq:gamma12:mat}
\bm{\gamma}_1  =\bm{\gamma}_2 = \left(
\begin{array}{cc}
\gamma_{11} & \gamma_{12}\\
\gamma_{12} & \gamma_{22}
\end{array}
\right) ,
\end{align}
where we have used the fact that, as mentioned above, the state is
symmetric by exchanging $\vec{x}_1$ and $\vec{x}_2$, so
$\bm{\gamma}_1=\bm{\gamma}_2$. The von-Neumann entropy can then be
written as
\begin{align}
S_1 = - \mathrm{Tr} \left[{\rho}_1 \log_2\left({\rho}_1\right)\right]\, ,
\end{align}
with similar expressions for $S_2$ and $S_{1,2}$. The quantity
$S_1=S_2$ is also called the entanglement entropy of the system, and
this allows us to evaluate $\mathcal{I}$ with \Eq{eq:calI:def}.

For a Gaussian state, the von-Neumann entropy is given
by~\cite{1999quant.ph.12067H}
\begin{align}
\label{eq:VonNeumann:Entropy:f}
  S({\rho})=\sum_{i=1}^nf(\sigma_i),
\end{align}
where the function $f(x)$ is defined for $x\geq 1$ by
\begin{align}
\label{eq:f(x):def}
  f(x)=\left(\frac{x+1}{2}\right)
  \log_2\left(\frac{x+1}{2}\right)
-\left(\frac{x-1}{2}\right)
\log_2\left(\frac{x-1}{2}\right),
\end{align}
and $\sigma_i$ are the symplectic eigenvalues of the covariance
matrix, that is to say the quantities $\sigma_i$ such that
$\mathrm{Sp}(\bm{J}^{(n)}\bm{\gamma})=\{i\sigma_1,-i\sigma_1,\cdots,
i\sigma_n,-i\sigma_n\}$. In this expression, we recall that
$\bm{J}^{(1)}=
  \begin{pmatrix}
    0 & 1 \\
    -1 & 0
  \end{pmatrix}$, and $\bm{J}^{(n)}$ is the $(2n\times 2n)$ block-diagonal
  matrix where each block corresponds to $\bm{J}^{(1)}$, and where $2 n$ is
  the dimension of phase space. 
  
In the present situation, the symplectic eigenvalues of the full
covariance matrix are given by
 \begin{align}
\label{eq:sigma:pm}
\sigma_\pm=\sqrt{\left(\gamma_{11}\pm\gamma_{13}\right)
  \left(\gamma_{22}\pm\gamma_{24}\right)
  -\left(\gamma_{12}\pm\gamma_{14}\right)^2}\, .
\end{align}
They allow one to rewrite \Eq{eq:det:gamma} as
$\det\gamma=\left(\sigma_+\sigma_-\right)^2$, which also follows from
the definition of the symplectic spectrum and the fact that
$\det[\bm{J}^{(n)}]=1$. For the reduced states, making use of
\Eq{eq:gamma12:mat}, one obtains a single symplectic eigenvalue, namely
\begin{align}
\label{eq:sigma1:def}
\sigma_{1} = \sqrt{\det\bm{\gamma}_1} =
\sqrt{\gamma_{11}\gamma_{22}-\gamma_{12}^2}\, .
\end{align}
Combining the above considerations, the mutual information is given by
\begin{align}
  \mathcal{I} (\vec{x}_1,\vec{x}_2)&= S\left({\rho}_1\right)
  + S\left({\rho}_2\right)
  - S\left({\rho}_{1,2}\right) 
  =  2 f\left(\sigma_{1}\right) - f\left(\sigma_+\right)
  - f\left(\sigma_-\right) \, .
 \label{eq:I:result}
\end{align}
Let us note that in the regime where the separation $d$ between
$\vec{x}_1$ and $\vec{x}_2$ is large, the cardinal sine suppression in
\Eq{eq:CorrelationMatrix:PowerSpectra} drives $\gamma_{13}$,
$\gamma_{14}$ and $\gamma_{24}$ to small values. In the limit where
they can be neglected, one has $\sigma_{\pm}\simeq \sigma_{1}$, which
leads to $\mathcal{I}\simeq 0$, hence more distant patches are less
correlated.  Another remark of interest is that the von-Neumann
entropy of a pure state is known to vanish,\footnote{This is because
  the density matrix of a pure state is idempotent, \ie
  $\rho^2=\rho$. Using the binomial expansion, this leads to
  $(\rho-\mathrm{Id})^k = (-1)^k (\mathrm{Id}-\rho)$, hence
  $\rho(\rho-\mathrm{Id})^k=0$ (where $k$ is a positive
  integer). Since $\log \rho$ is defined as a Taylor series by
  $\log \rho = \sum_{k=1}^\infty (-1)^{k+1} (\rho-\mathrm{Id})^k/k$,
  one has $\rho\log \rho=0$, so pure states have indeed vanishing
  entropy.} so the mutual information between two subsystems of a
pure state simply corresponds to twice the entanglement entropy.

From~\Eqs{eq:VonNeumann:Entropy:f} and~\eqref{eq:f(x):def}, one can
see that a state with vanishing von-Neumann entropy is one for which
the symplectic eigenvalues all equal one. Here, and as will be made
more explicit in \Sec{sec:Cosmo}, $\sigma_{\pm}$ are not equal to one
in general, denoting the fact that we are not dealing with a pure
state.  This may seem surprising, since in this work we consider a
single scalar field $\phi(\vec{x})$, isolated from any environmental
degrees of freedom, and which can therefore be placed in a pure
state. The reason is the following. If the field is placed on a
homogeneous background, its quantum state is separable in Fourier
space, which means that there are no correlations between the Fourier
subspaces $\vec{k}_1$ and $\vec{k}_2$ if $\vec{k_1}\neq \pm
\vec{k}_2$. This was shown explicitly around
\Eq{PowerSpectrum:interm2}. The reduced state within each Fourier
subspace $\pm\vec{k}$ may therefore be pure, and within each Fourier
subspace, one can study the presence of (quantum) correlations between
the sectors $\vec{k}$ and $-\vec{k}$ as subsectors of a pure
state~\cite{Martin:2015qta}. In real space however, the field
generally features non-trivial correlations, see
\Eq{eq:PowerSpectrum:RealSpace}. Therefore, when restricting one's
attention to its (coarse-grained) configurations at locations
$\vec{x}_1$ and $\vec{x}_2$, one implicitly traces over its
configuration at all other locations (to which the configurations at
$\vec{x}_1$ and $\vec{x}_2$ are entangled), which leads to a non-pure
bipartite system. In general, this effective ``self-decoherence'' can
be measured with the purity parameter~\cite{PhysRevD.24.1516,
  PhysRevD.26.1862, Joos:1984uk,Colas:2021llj}
\begin{align}
\label{eq:purity:def}
\mathfrak{p} = \mathrm{Tr}(\rho^2)=\frac{1}{\sqrt{\det\bm{\gamma}}}
= \frac{1}{\prod_i \sigma_i}\, .
\end{align}
The last expression may be used to characterise either the full system
$\rho_{1,2}$ or the reduced systems $\rho_{1}=\rho_2$, by considering
the relevant symplectic eigenvalues in each case. Pure states have
$\mathfrak{p}=1$, while decohered states are such that $0\leq
\mathfrak{p}<1$. Decoherence usually leads to a suppression in the
amount of quantum correlations~\cite{2013IJMPB..2745042B}. This already hints to the fact that even if large
entanglement is present in Fourier space, real-space measurements may
feature less quantum correlations than those encountered in Fourier
space.
\subsection{Quantum discord}
\label{subsec:discord}
One way to measure the ``quantumness'' of the correlations between the
field configurations at locations $\vec{x}_1$ and $\vec{x}_2$ is via
quantum discord, which we now introduce. This is first done at the
level of classical random variables, in the same language as the one
employed at the beginning of \Sec{sec:MutualInformation}. Upon
denoting $p_{i\vert j}$ the conditional probability to find the field
in configuration $a_i$ at location $\vec{x}_1$ knowing that it is in
configuration $b_j$ at $\vec{x}_2$, Baye's theorem leads to $p_{i,j} =
p_j p_{i\vert j}$. When plugging this relation into the
definition~\eqref{eq:calI:def} of mutual information, one obtains
$\mathcal{I} = -\sum_i p_i \log_2(p_i)-\sum_j p_j \log_2(p_j) +
\sum_{i,j} p_jp_{i\vert j} [\log_2(p_j)+\log_2(p_{i\vert j})] =
-\sum_i p_i \log_2(p_i) + \sum_{i,j} p_jp_{i\vert j} \log_2(p_{i\vert
  j})$ where we have used that $\sum_i p_{i\vert j}=1$. This suggests
introducing the quantity
\begin{align}
\label{eq:conditional:entropy}
S_{1\vert 2} = - \sum_{j }p_j  \sum_i p_{i\vert j} \log_2(p_{i\vert j})\, ,
\end{align}
which stands for the conditional entropy contained in the field
configuration at $\vec{x}_1$ after finding the field in configuration
$b_j$ at $\vec{x}_2$, averaged over all possible configurations at
$\vec{x}_2$. The above calculation thus gives rise to an alternative
expression for mutual information, namely
\begin{align}
\label{eq:def:J}
\mathcal{J}(\vec{x}_1,\vec{x}_2)
 = S_1-S_{1\vert 2}\, .
\end{align}
These considerations show that, in classical systems,
$\mathcal{I}=\mathcal{J}$.

This is however not necessarily the case in quantum systems, and the
fact that $\mathcal{I}-\mathcal{J}$ vanishes for classical systems
only can be used to define a criterion for the presence of quantum
correlations. First, one needs to translate the conditional
entropy~\eqref{eq:conditional:entropy} into the quantum formalism.

To this end, let us introduce ${\Pi}_j$, a complete set of projectors
on the field configurations at $\vec{x}_2$, and denote by $\vert
b_j\rangle$ the quantum states on which they project. One thus has
${\Pi}_j = {\setI}_1\otimes \vert b_j \rangle \langle b_j \vert$. Let
us note that such complete sets of projectors ${\Pi}_j$ (or
equivalently, of states $\vert b_j \rangle$) are not unique (for a
spin particle for instance, one can consider $\vert +
\rangle_{\vec{e}}$ and $\vert - \rangle_{\vec{e}}$ along any unit
vector $\vec{e}$), and this fact will be dealt with below.  The
probability to find the field at $\vec{x}_2$ in the state $b_j$ is
given by $p_j = \mathrm{Tr}({\rho} {\Pi}_j)$, and a measurement of the
field at $\vec{x}_2$ that returns the result $b_j$ projects the state
into ${\rho} \to {\Pi}_j{\rho} {\Pi}_j/p_j $. This leads us to
introducing
\begin{align}
  {\rho}_{1\vert \hat{\Pi}_i} = \mathrm{Tr}_2
  \left(\frac{ {\Pi}_j{\rho} {\Pi}_j}{p_j}\right),
\end{align}
which is the state of the field at $\vec{x}_1$ after measuring its
configuration at $\vec{x}_2$ and finding $b_j$ as a result of the
measurement. The conditional entropy can thus be written as
\begin{align}
S_{1\vert 2} = \sum_j p_j S\left(\hat{\rho}_{1\vert \hat{\Pi}_i} \right) .
\end{align}
This is the analogue of \Eq{eq:conditional:entropy}, and these
formulas then allow one to evaluate $\mathcal{J}$ with
\Eq{eq:def:J}. Quantum discord is finally defined as
\begin{align}
\label{eq:discord:def}
  \mathcal{D}(\vec{x}_1,\vec{x}_2)= \min_{\{\hat{\Pi}_i\}}\left[{\cal I}(\vec{x}_1,\vec{x}_2)-{\cal J}(\vec{x}_1,\vec{x}_2)\right]\, ,
\end{align}
where minimisation is performed over all possible complete sets of
projectors, in order to ensure that a non-vanishing discord signals
the presence of genuine quantum correlations for any projection basis.

A generic calculation of quantum discord for Gaussian states is
presented in \Refa{2010PhRvL.105c0501A}. In this article we only state
the result in terms of the covariance matrix $\bm{\gamma}$, but a
detailed derivation of the formulas below can be found in this
reference. Let us first denote by
$\bm{\gamma}_{1\mhyphen 2}$ the off-diagonal block of the covariance
matrix,
\begin{align}
\bm{\gamma}_{1\mhyphen 2} = \left(
\begin{array}{cc}
\gamma_{13} & \gamma_{14}\\
\gamma_{14} & \gamma_{24}
\end{array}
\right),
\end{align}
such that the covariance matrix can be written in the block form as
$
\bm{\gamma} = \left(
\begin{array}{cc}
\bm{\gamma}_{1} & \bm{\gamma}_{1\mhyphen 2}\\
\bm{\gamma}_{1\mhyphen 2} & \bm{\gamma}_{1}
\end{array}
\right) .
$
Similarly to \Eq{eq:sigma1:def}, we introduce
\begin{align}
\label{eq:sigma12:def}
\sigma_{1\mhyphen 2} = \sqrt{\det\bm{\gamma}_{1\mhyphen 2}}
= \sqrt{\gamma_{13}\gamma_{24}-\gamma_{14}^2}\, ,
\end{align}
where one should note that $\sigma_{1\mhyphen 2}^2$ may be positive or
negative.  One can show that $\sigma_+$, $\sigma_-$, $\sigma_1$ and
$\sigma_{1\mhyphen 2}$ are all local-symplectic invariants, which
means that they are invariant under canonical transformations that act
on each sector $\vec{x}_1$ and $\vec{x}_2$ separately (\ie
transformations represented by block-diagonal symplectic
matrices). Quantum discord, which is also a local-symplectic
invariant, can therefore be written in terms of these quantities only, and
after extremisation over the set of projectors appearing in
\Eq{eq:discord:def} one has~\cite{2010PhRvL.105c0501A}
\begin{align}
\label{eq:J}
\mathcal{J}(\vec{x}_1,\vec{x}_2) &= f\left(\sigma_1\right)
-f\left(\sqrt{E}\right),
\end{align}
with
\begin{align}
  \label{eq:E}
E = 
\begin{cases}
  \displaystyle
  \frac{1}{(\sigma_1^2-1)^2}\biggl\{2\sigma_{1\mhyphen 2}^4
  +\left(\sigma_1^2-1\right)
    \left(\sigma_+^2\sigma_-^2-\sigma_1^2\right)
    +2\left\vert
    \sigma_{1\mhyphen 2}^2\right\vert \bigl[\sigma_{1\mhyphen 2}^4
      +\left(\sigma_1^2-1\right)
\\ \hspace{1.8cm}\times
      \left(\sigma_+^2\sigma_-^2-\sigma_1^2\right)\bigr]^{1/2}\biggr\},
\\ \\
\displaystyle
\frac{1}{2\sigma_1^2}\left[\sigma_1^4-\sigma_{1\mhyphen 2}^4+\sigma_+^2\sigma_-^2
  -\sqrt{\sigma_{1\mhyphen 2}^8+\left(\sigma_1^4-\sigma_+^2\sigma_-^2\right)^2
    -2\sigma_{1\mhyphen 2}^4\left(\sigma_1^4+\sigma_+^2\sigma_-^2\right)}\right]\, ,
\end{cases}
\end{align}
where the first equality applies if
$\left(1+\sigma_1^2\right) \sigma_{1\mhyphen
  2}^4\left(\sigma_1^2+\sigma_+^2\sigma_-^2\right)
-\left(\sigma_1^4-\sigma_+^2\sigma_-^2\right)^2\geq 0$ and the second
one otherwise. As for the mutual information $\mathcal{I}$, one can
check that in the limit where the configurations of the field at
$\vec{x}_1$ and $\vec{x}_2$ are uncorrelated, $\gamma_{13}$,
$\gamma_{14}$ and $\gamma_{24}$ can be neglected, hence
$\sigma_{1\mhyphen 2}$ can be neglected and one obtains
$\mathcal{J}=0$ [where one also has to use that
  $\sigma_+^2\sigma_-^2<\sigma_1^4$, which directly follow from
  \Eqs{eq:sigma:pm} and~\eqref{eq:sigma1:def}].

The above considerations provide all necessary formulas to explicitly
compute the mutual information and the quantum discord between the
field configurations at $\vec{x}_1$ and $\vec{x}_2$ from the knowledge
of the power spectra.
\section{Application to cosmological perturbations}
\label{sec:Cosmo}

We now consider the main question investigated in this article and use the formalism presented above to study cosmological
perturbations. This will lead us to establishing important
results for the detectability of quantum correlations in cosmological
measurements, shedding light on our ability to prove or disprove
their quantum-mechanical origin.

Let us therefore consider the case of a homogeneous and isotropic cosmology,
described by the flat Friedmann-Lema\^itre-Robertson-Walker metric
\begin{align}
\dd s^2 = -\dd t^2 + a^2(t) \dd \vec{x}^2\, ,
\end{align}
where $a(t)$ is the scale factor.  Density fluctuations are described
by the curvature perturbation
$\zeta(\vec{x})$~\cite{Mukhanov:1981xt,Kodama:1985bj}, which is a
diffeomorphism-invariant combination of scalar fluctuations of the
metric components and of the matter sector. On large scales, it is
directly proportional to the temperature anisotropies measured on the
CMB~\cite{Akrami:2018vks}. It can also be described by the
Mukhanov-Sasaki variable $v(\vec{x})=z \zeta(\vec{x}) $, where
$z\equiv a\sqrt{2\epsilon_1}\Mp/\cs$ where $\cs$ is the speed of sound
($\cs=1$ for a scalar field) and $\epsilon_1\equiv - \dot{H}/H^2$ is
the first Hubble-flow parameter~\cite{Schwarz:2001vv,Leach:2002ar},
with $H=\dot{a}/a$ the Hubble parameter and a dot denotes derivation
with respect to cosmic time $t$. In Fourier space, the conjugate
momentum to $v_{\vec{k}}$ is ${p}_{\vec{ k}}\equiv {v}_{\vec{ k}}' -
(z'/z) v_{\vec{k}}$, where a prime denotes derivation with respect to
the conformal time $\eta$ defined via $\dd t= a \dd\eta$. Each mode
behaves as a parametric oscillator, with equation of motion
\begin{align}
\label{eq:Mukhanov:Sasaki}
v_{\vec{k}}''+ \left(\cs^2k^2-\frac{z''}{z}\right) v_{\vec{k}}=0\, .
\end{align}
The standard cosmological scenario starts with a phase of accelerated
expansion ($\ddot{a}>0$) called cosmic inflation, during which quantum
vacuum fluctuations are gravitationally amplified and stretched to
astrophysical distances, which seeds the cosmological structures we
later observe. In what follows, we study the quantum correlations
contained in cosmological perturbations during this early epoch of
inflation, and during the subsequent era where the universe is
dominated by radiation.
\subsection{Inflationary era}
\label{sec:Inflation}

\subsubsection{Calculation of the covariance matrix}
\label{subsubsec:covinf}

During inflation, measurements of the CMB constrain the expansion of
the universe to proceed close to the de-Sitter regime where
$a=-1/(H\eta)$. The Mukhanov-Sasaki
equation~\eqref{eq:Mukhanov:Sasaki} thus reads
$v_{\vec{k}}''+(k^2-2/\eta^2)v_{\vec{k}}=0$, the solution of which is
given by
\begin{align}
\label{eq:infl:def}
v_{\vec{k}}=\frac{\ee^{-i k \eta}}{\sqrt{2 k}}\left(1-\frac{i}{k\eta}\right) . 
\end{align}
Here, the mode function has been normalised to the Bunch-Davies
vacuum~\cite{Bunch:1978yq}, \ie the integration constants are set such
that in the asymptotic past, $v_{\vec{k}} \propto \ee^{-i k
  \eta}/\sqrt{2k}$ matches the Minkowski vacuum. The momentum
conjugated to $v_{\vec{k}}$ is
\begin{align}
\label{eq:pk:def}
p_{\vec{k}}  = -i \sqrt{\frac{k}{2}}\ee^{-ik\eta}.
\end{align}
Hereafter we make the identification $\phi(\vec{x})=v(\vec{x})$ and
$\pi(\vec{x}) = p(\vec{x})$, but one should recall that since the
quantities introduced in \Sec{sec:MutualInformation:Discord} are
local-symplectic invariants, the following calculations do not depend
on the choice of canonical variables. In other words, one may perform
any canonical transformation (for instance, describe the system with
the curvature fluctuation $\zeta$ and its conjugate momentum) without
affecting the result. The reason why we choose to work with the
Mukhanov-Sasaki variable is one of convenience, and \Eqs{eq:infl:def}
and~\eqref{eq:pk:def} lead to the reduced power spectra
\begin{align}
\label{eq:PowerSpectra:deSitter}
\calP_{vv}(k) = \frac{1+k^2\eta^2}{4\pi^2\eta^2}\, ,\qquad
\calP_{pp}(k) = \frac{k^4}{4\pi^2}\, ,\qquad
\calP_{vp}(k) = \frac{k^2}{4\pi^2\eta}\, .
\end{align}
One can check that $(2\pi^2/k^3)^2\left[\calP_{vv}(k)
    \calP_{pp}(k) - \calP_{vp}^2(k)\right] = 1/4$, which confirms the
remark made at the end of \Sec{sec:MutualInformation} that each
Fourier mode is placed in a pure state. Indeed, recalling
\Eq{eq:Covariance:matrix:def}, it implies that the determinant of the
covariance matrix describing a given Fourier mode is one, so the
purity equals one too, see \Eq{eq:purity:def}.

\begin{figure}[t]
\begin{center}
\includegraphics[width=0.6\textwidth]{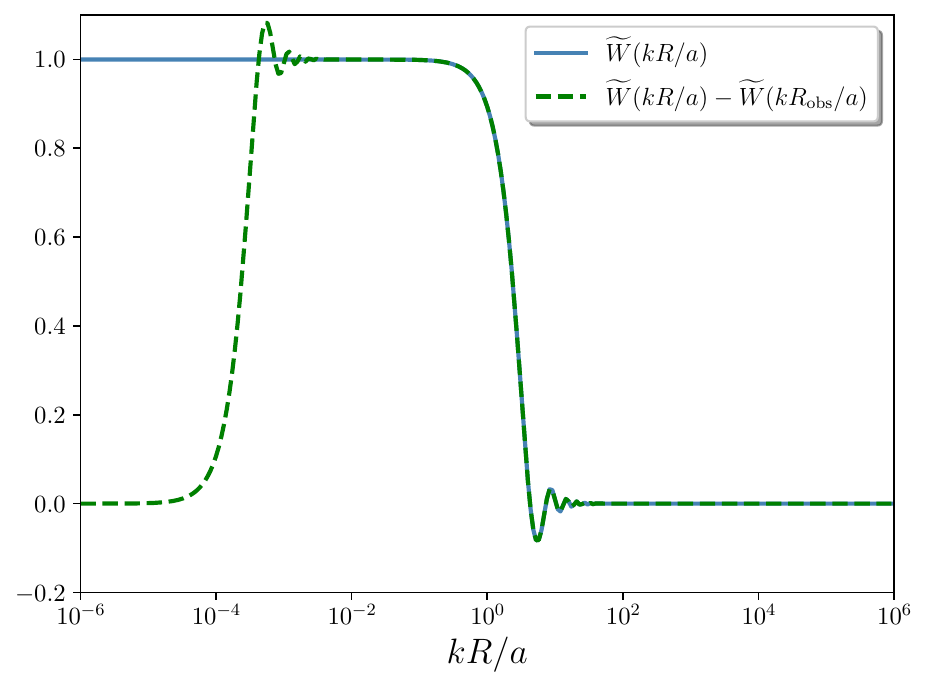}
\caption{Window
  function~\eqref{eq:Fourier:transform:Heaviside:Improved} as a
  function of $k R/a$ (solid blue line), and effective window function
  appearing in \Eq{eq:local:IR:removal} once the finite size of the
  observable universe is taken into account (dashed green line). The
  parameter $\delta$ has been set to $\delta=0.1$, and $R/R_{\mathrm{obs}}=10^{-4}$.}
\label{fig:WindowFunction}
\end{center}
\end{figure}
When computing the covariance matrix via
\Eq{eq:CorrelationMatrix:PowerSpectra}, one may note that some of the
integrals over $k$ are IR-divergent, \ie they diverge when $k\to
0$. Indeed, as noticed below
\Eq{eq:Fourier:transform:Heaviside:Improved}, when $k\ll a/R$,
$\widetilde{W}(kR/a)\simeq 1$, and one also has
  $\mathrm{sinc}(kd/a)\simeq 1$ as soon as $k\ll a/d$. From
\Eq{eq:PowerSpectra:deSitter}, one thus finds that $\gamma_{11}$ and
$\gamma_{13}$ are logarithmically divergent. The reason why this
divergence does not appear in practice is the following. Any local
observer only has access to a finite region of the universe, and we
let $R_{\mathrm{obs}}$ denote the size of the observable universe. In
terms of more usual cosmological parameters, it can be written as
\begin{align}
    R_\mathrm{obs} = \frac{\ee^{N_\mathrm{inf}}}{H}\, ,
\end{align}
where $N_{\mathrm{inf}}$ is the number of \efolds~spent outside the
Hubble radius $H^{-1}$ by the largest observable scales (it is
typically of order $50$ at the end of inflation), and $H$ denotes the
almost-constant value of the Hubble parameter during inflation. In
practice, ``fluctuations'' are perceived as deviations of the fields
$\bm{z}_R(\vec{x})$ from the mean value measured inside the observable
patch, \ie one only has access to
\begin{align} \delta
  \bm{z}_R(\vec{x}) = \bm{z}_R(\vec{x}) -
  \bm{z}_{R_{\mathrm{obs}}}(\vec{x}_0),
\end{align}
where $\vec{x}_0$ represents the location of the observer. Making
use of \Eq{eq:def:tilde:W}, this gives rise to
\begin{align}
\label{eq:local:IR:removal}
\delta\bm{z}_R(\vec{k}) = \bm{z}(\vec{k}) \left[\widetilde{W}
  \left(\frac{kR}{a}\right)-\widetilde{W}
  \left(\frac{kR_\mathrm{obs}}{a}\right) \right] .
\end{align}
This means that, once the finite size of the observable universe is
taken into account, the window function becomes $\widetilde{W}(kR/a)
\to \widetilde{W}(kR/a) - \widetilde{W}(kR_{\mathrm{obs}}/a)$. Because
of the generic properties of the function $\widetilde{W}$ discussed
below \Eq{eq:def:tilde:W}, when $k\gg a/R_\mathrm{obs}$, \ie for
wavenumbers inside the observable patch, this does not modify the
window function substantially. However, when $k\ll
a/R_{\mathrm{obs}}$, the two terms in the effective window function
cancel out each other, which implies that unobservable modes, \ie those
above the observed region, are filtered out. This is confirmed by
\Fig{fig:WindowFunction} where both $\widetilde{W}(kR/a)$ and
$\widetilde{W}(kR/a) - \widetilde{W}(kR_{\mathrm{obs}}/a)$ are
displayed as a function of $kR/a$. The effective window function thus
selects out modes such that $a/R_{\mathrm{obs}}<k<a/R$.

Hereafter, the finite size of the observable universe is taken into
account by simply adding a lower bound to all $k$-integrals at
$k=a/R_{\mathrm{obs}}$, \ie the effective window function one
considers is $\widetilde{W}(kR/a) \to \widetilde{W}(kR/a)
\theta(kR/a- R/R_{\mathrm{obs}})$. The reason is that the details
of the terms coming from this lower bound only play a minor role in
the formulas derived below, especially when considering the relevant
limit $R_{\mathrm{obs}}\gg R$. In practice, this solves the IR
divergence mentioned above.

\begin{figure}[t]
\begin{center}
\includegraphics[width=0.89\textwidth]{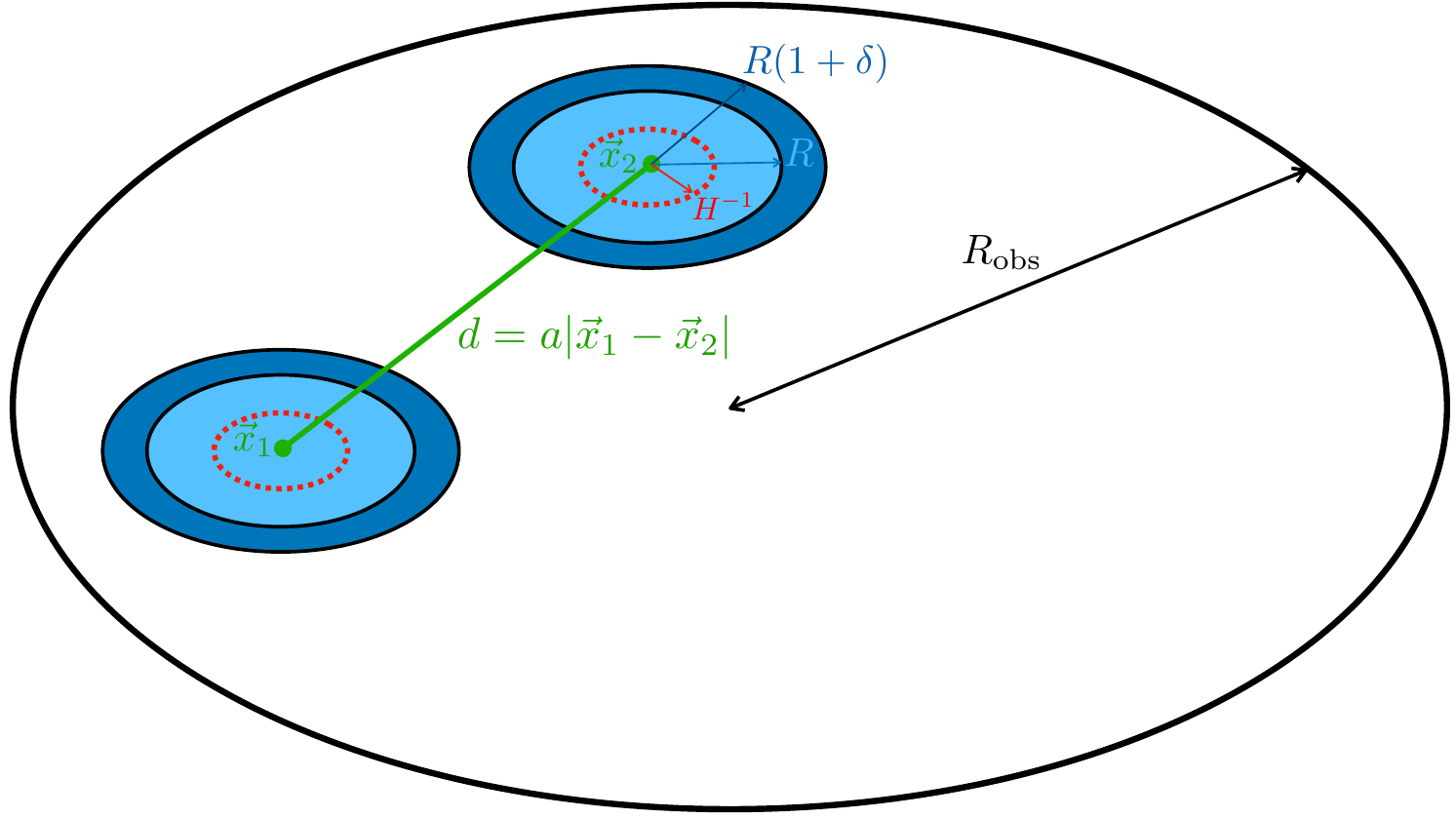}
\caption{Sketch of the different scales involved in the calculation of
  entanglement in real space. The two coarse-grained patches, located at
  $\vec{x}_1$ and $\vec{x}_2$, have a size $R$ [and $R(1+\delta)$ if one includes the full extent of the window function, where $\delta\lesssim 1$]. They are comprised within the observable region of the universe, depicted with the black ellipse, the size of which is denoted $R_{\mathrm{obs}}$. The distance between the patches is $d=a\vert\vec{x}_1-\vec{x}_2\vert$, and for the two patches not to overlap, one must have $d>2R(1+\delta)$. The Hubble radius is displayed with the red ellipse. In the situation depicted in this figure, it is smaller than the size of the patch, \ie $HR>1$,  but all possible configurations are studied in the present work.}
\label{fig:sketchscale}
\end{center}
\end{figure}

With \Eq{eq:PowerSpectra:deSitter} for the power spectra, the entries
of the covariance matrix given in
\Eq{eq:CorrelationMatrix:PowerSpectra} read
\begin{align}
  \label{eq:Cov:Inflation:exact}
  \gamma_{11}&=\frac{2\lambda^2}{3\pi{G(\delta)}}(HR)^2
  \left(\frac{R}{a}\right)
  \left[{\cal K}(\beta,-1,\delta)+\frac{1}{(HR)^2}
    {\cal K}(\beta,1,\delta)\right],
  \\
  \label{eq:gamma12}
  \gamma_{12}&=-\frac{2}{3\pi{G(\delta)}}(HR)\, {\cal K}(\beta,1,\delta),
  \qquad
  \gamma_{22}=\frac{2}{3\pi \lambda^2{G(\delta)}}\left(\frac{R}{a}\right)^{-1}
        {\cal K}(\beta,3,\delta),
        \\
        \label{eq:gamma13}
        \gamma_{13}&=\frac{2\lambda^2}{3\pi{G(\delta)}}(HR)^2
        \left(\frac{R}{a}\right)
        \left[{\cal L}(\beta,-1,\delta,\alpha)+\frac{1}{(HR)^2}
          {\cal L}(\beta,1,\delta,\alpha)\right],      
  \\
  \label{eq:gamma14}
\gamma_{14}&=-\frac{2}{3\pi{G(\delta)}}(HR)\, {\cal L}(\beta,1,\delta,\alpha),
\qquad
\gamma_{24}=\frac{2}{3\pi \lambda^2{G(\delta)}}\left(\frac{R}{a}\right)^{-1}
      {\cal L}(\beta,3,\delta,\alpha).
\end{align}
In these expressions, we have introduced a few relevant parameters and
useful notations that we now describe (see also \Fig{fig:sketchscale}). The parameter $\alpha$ denotes
the distance between the two patches in units of $R$,
\begin{align}
\label{eq:aplha:cond}
\alpha \equiv \frac{a \vert \vec{x}_1 - \vec{x}_2\vert}{R}
> 2\left(1+\delta\right)\, ,
\end{align}
where the lower bound comes from \Eq{eq:lowerbound:d}. The parameter
$\beta$ corresponds to the lower bound imposed on $kR/a$ in order to
take into account the finite size of the observable universe, \ie
\begin{align}
\label{eq:beta:def}
\beta=\frac{R}{R_\mathrm{obs}}<1\, .
\end{align}
The condition $\beta<1$ comes from the fact that one cannot
coarse-grain over distances larger than those observable, but the
relevant regime really is $\beta\ll 1$ (for instance, in CMB
measurements, $\beta$ is of the order of the inverse of the maximal
multipolar moment $\ell_\umax\sim 2500$, and is even much smaller for
measurements of the large-scale structure performed at smaller
redshifts). 
Finally, since
the power spectra~\eqref{eq:PowerSpectra:deSitter} only feature
power-law functions of the wavenumber, the covariance matrix only
involves integrals of the type
\begin{align}
\label{eq:KLM:def}
      {\cal K}(\beta,\mu,\delta)&\equiv
      \int_\beta^{\infty}z^{\mu}\, \widetilde{W}^2(z) \, \dd z,
      \\
      \label{eq:defL}
  {\cal L}(\beta,\mu,\delta,\rho) &\equiv
  \int_\beta^{\infty}z^{\mu}\, \widetilde{W}^2(z)\, \mathrm{sinc} (\rho z)
  \, \dd z ,
  \\
  \label{eq:defM}
   {\cal M}(\beta,\mu,\delta,\rho) &\equiv
   \int_\beta^{\infty}z^{\mu}\, \widetilde{W}^2(z)
   \cos (\rho z)\,  \dd z ,
\end{align}
where the integral $\mathcal{M}$ has also been introduced since it
will be needed in the calculation of \Sec{sec:Radiation} where the
covariance matrix is obtained in the radiation-dominated era. Since
the window function~\eqref{eq:Fourier:transform:Heaviside:Improved}
involves trigonometric and power-law functions of $kR/a$, these three
integrals can be expressed solely in terms of the cosine integral function, see
\App{app:Integrals} for details and explicit formulas.

The above expressions~(\ref{eq:Cov:Inflation:exact}),
(\ref{eq:gamma12}), (\ref{eq:gamma13}) and~(\ref{eq:gamma14}) of the
covariance matrix are quite involved and, as a consequence, it is
interesting to derive analytical approximations for the relevant
physical quantities. One can check that the factors $\lambda$ and
$R/a$ appearing in \Eq{eq:Cov:Inflation:exact} cancel out when
computing the symplectic values,\footnote{The fact that $\lambda$
  cancels out is expected from the above remark that it simply
  corresponds to a local canonical redefinition of the phase-space
  variables, and the independence on $R/a$ follows from the fact that
  it can be re-absorbed by a rescaling of $\lambda\to
  \lambda\sqrt{a/R}$.\label{footnote:lambda}} so their values depend
only on four parameters, namely $\alpha$, $\beta$, $\delta$ and
$HR$. As explained around \Eq{eq:aplha:cond}, one must have
$\alpha>2(1+\delta)$, so it is interesting to consider the limit
$\alpha\gg 1$ which corresponds to the situation where the two
observed patches are well separated, \ie $d\gg R$. The parameter
$\beta$, defined in \Eq{eq:beta:def}, corresponds to the ratio between
the size of the observed patches and the size of the entire observable
universe, which is why, as already mentioned, the regime $\beta\ll 1$ is of
interest. Finally, $\delta$ is a parameter that describes the edge of
the window function in real space. It is smaller than one for
experiments with a sharp filtering device, so for convenience we
expand our results in $\delta$ too, although the precise value of
$\delta$ is of little importance in what follows, as long as it
remains of order one or smaller. Let us note that
$\alpha\beta=d/R_{\mathrm{obs}}$, so $\beta\ll 1/\alpha$ when the
observed patches are well within the observable universe. This is why
one should first expand in $\beta\ll 1$, and then in $\alpha\gg 1$
before finally expanding in $\delta$. When doing so, in
\App{app:Integrals}, approximate expressions for the integrals
$\mathcal{K}$ and $\mathcal{L}$ are obtained. Plugging the result into
\Eq{eq:Cov:Inflation:exact}, (\ref{eq:gamma12}), (\ref{eq:gamma13})
and~(\ref{eq:gamma14}), one obtains
\begin{align}  
\label{eq:Cov:Inflation:appr}
\gamma_{11} \simeq &\frac{2 (HR)^2}{3\pi}
\biggl[-\left(1+2\delta\right)\ln(2\beta)+\frac{7}{4}
-\gamma_\mathrm{E}+\left(3-2\gamma_\mathrm{E}\right)
\delta+\frac{\beta^2}{10}(1+3\delta)
+\frac{9(1+\delta)}{4(HR)^2}\biggr],
\\
\label{eq:covapprox12}
\gamma_{12} \simeq &
-\frac{3}{2\pi}HR\left[1+\delta-\frac{2}{9}\beta^2\left(1+2\delta\right)\right],
\\
\label{eq:covapprox22}
\gamma_{22}\simeq &
\frac{3}{2\pi}\left[1+\delta -2\ln \frac{\delta}{2}
  -\frac{\beta^4}{9}(1+2\delta)\right],
\\
\label{eq:covapprox13}
\gamma_{13}
\simeq &\frac{2\left(HR\right)^2}{3\pi} \left\lbrace
\left[-\ln\left(\alpha\beta\right)+1-\gamma_\mathrm{E}\right](1+2\delta)
{-\frac{1+3\delta}{5\alpha^2}+\frac{\alpha^2\beta^2}{12}(1+2\delta)
+\frac{1+2\delta}{\alpha^2(HR)^2}}\right\rbrace,
\\
\label{eq:covapprox14}
\gamma_{14} \simeq &-\frac{2}{3\pi}\frac{HR}{\alpha^2}
\left[1+2\delta+{\frac{2}{5\alpha^2}(1+3\delta)
-\frac{\alpha^2\beta^2}{2}(1+2\delta)}\right],
\\
\label{eq:covapprox24}
\gamma_{24} \simeq &
-\frac{4}{3\pi\alpha^4}\left[1+2\delta
  +{\frac{12}{5\alpha^2}(1+3\delta)+\frac{\alpha^4\beta^4}{8}
    (1+2\delta)}\right],
\end{align}
where we have set $\lambda=\sqrt{a/R}$ for convenience, see
footnote~\ref{footnote:lambda}. On top of the parameters already
mentioned, these expressions also involve the combination $HR$, and different regimes for the value of that parameter will have to be distinguished below.

\subsubsection{De Sitter mutual information}
\label{subsubsec:mutualdS}

Having established the relevant expressions for the covariance matrix, one can now determine the symplectic values
$\sigma_+$, $\sigma_-$, $\sigma_1$ and $\sigma_{1\mhyphen 2}$, and the
mutual information. Since we are far from resolving the
Hubble radius during inflation (that would imply to measure the CMB up
to multipoles $\ell\sim \ee^{50}$), the relevant limit is $HR\gg
1$. Plugging \Eqs{eq:Cov:Inflation:appr}-\eqref{eq:covapprox24} into \Eqs{eq:sigma:pm},
\eqref{eq:sigma1:def} and~\eqref{eq:sigma12:def}, one obtains
\begin{align}
\label{eq:sigma:infl:appr}
\sigma_+^2\simeq &
\left(\frac{HR}{\pi}\right)^2\left
\{\left[-\left(1+2\delta\right)\ln(2\alpha\beta^2)
  +\frac{11}{4}-2\gamma_\mathrm{E}\right]
\left(1-2\ln \frac{\delta}{2}\right)-\frac{9}{4}\right\}
\\
\label{eq:sigmaminf}
\sigma_-^2 \simeq &
\left(\frac{HR}{\pi}\right)^2\left\{\left[(1+2\delta)
  \ln \frac{\alpha}{2}+\frac{3}{4}\right]
\left(1-2\ln \frac{\delta}{2}\right)-\frac{9}{4}\right\}
\\
\label{eq:sigma1inf}
\sigma_1^2 \simeq &
\left(\frac{HR}{\pi}\right)^2\left\{ \left(1-2\ln \frac{\delta}{2}\right)
\left[-\left(1+2\delta\right)\ln(2\beta)+\frac{7}{4}-\gamma_\mathrm{E}\right]
-\frac{9}{4} \right\}
\\
\label{eq:sigma12inf}
\sigma_{1\mhyphen 2}^2\simeq & \frac{8(HR)^2}{9\pi^2\alpha^4}
\left[\gamma_{\mathrm{E}}-\frac{3}{2}+\left(4\gamma_{\mathrm{E}}-5\right)\delta
  +\left(1+4\delta\right)\ln(\alpha \beta)\right]\, .
\end{align}
One can check that $\sigma_+^2$, $\sigma_-^2$ and $\sigma_1^2$ are all
positive under the conditions where this limit has been taken, which
is a good consistency check (recall that the sign of
$\sigma_{1\mhyphen 2}^2$ is not constrained).
%
%

\begin{figure}[t]
\begin{center}
\includegraphics[width=0.8\textwidth]{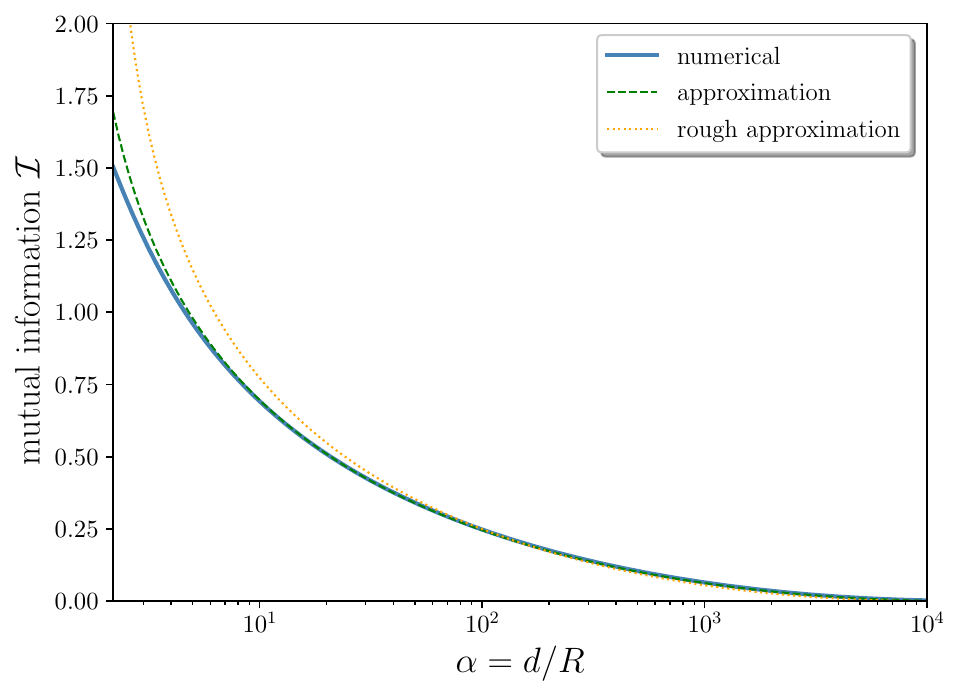}
\caption{Mutual information during cosmic inflation for
  $\beta=10^{-4}$, $HR=10^8$ and $\delta=0.1$, as a function of
  $\alpha$, which is varied in the physical range $\alpha\in
  [2(1+\delta),1/\beta]$, corresponding to
  $2R(1+\delta)<d<R_{\mathrm{obs}}$. The solid blue curve corresponds
  to a full numerical calculation, the green dashed curve stands for
  the approximated formula~\eqref{eq:calI:appr:inflation}, while the
  dotted orange curve displays the rough
  approximation~\eqref{eq:calI:appr:rough:inflation}.}
\label{fig:I:infl}
\end{center}
\end{figure}

Let us now compute the mutual information $\mathcal{I}$.  One can see
that $\sigma_+$, $\sigma_-$ and $\sigma_1$ are all large and of the
same order $HR$.  This means that the function $f$ appearing in
\Eq{eq:I:result}, and defined in \Eq{eq:f(x):def}, needs to be
evaluated with large arguments. Using that $f(x)\simeq
1/\ln 2 +\log_2(x/2)$ when $x\gg 1$, \Eq{eq:I:result} gives rise to
$\mathcal{I}\simeq
2\log_2(\sigma_{1})-\log_2(\sigma_+)-\log_2(\sigma_-)$, which leads to
\begin{align}
\label{eq:calI:appr:inflation}
\mathcal{I} (\vec{x}_1,\vec{x}_2)\simeq &
\log_2\left\lbrace \left(1-2\ln \frac{\delta}{2}\right)
\left[-\left(1+2\delta\right)\ln(2\beta)+\frac{7}{4}-\gamma_\mathrm{E}\right]
-\frac{9}{4} \right\rbrace
\nonumber
\\ &
-\frac{1}{2}\log_2\left\lbrace\left[-\left(1+2\delta\right)\ln(2\alpha\beta^2)
  +\frac{11}{4}-2\gamma_\mathrm{E}\right]\left(1-2\ln \frac{\delta}{2}\right)
-\frac{9}{4}\right\rbrace
\nonumber
\\ &
-\frac{1}{2}\log_2\left\lbrace\left[(1+2\delta)\ln \frac{\alpha}{2}
  +\frac{3}{4}\right]\left(1-2\ln \frac{\delta}{2}\right)
-\frac{9}{4}\right\rbrace\, .
\end{align}
One can see that the dependence on the parameters of the problem is
very mild, and that the result does not depend on $HR$ at that
order. In the limit where the logarithmic terms dominate over the
constant terms, a rough version of the above formula is given by
\begin{align}
\label{eq:calI:appr:rough:inflation}
\mathcal{I}(\vec{x}_1,\vec{x}_2)\sim
\frac{1}{2}\log_2\left[\frac{-\ln^2(2\beta)}
  {\ln(\alpha/2)\ln(2\alpha\beta^2)}\right] \, ,
\end{align}
which makes this mild dependence even more explicit (parameters only
appear through logarithms of logarithms), and in which $\delta$ no
longer appears. These approximated formulas are compared with a full
numerical calculation of the mutual information in
\Fig{fig:I:infl}. One can see that, in the regime $1\ll \alpha\ll
1/\beta$, the approximations indeed provide an excellent fit to the
exact result. Even if \Eq{eq:calI:appr:inflation} is more accurate
than \Eq{eq:calI:appr:rough:inflation} as expected,
\Eq{eq:calI:appr:rough:inflation} still provides a correct estimate
sufficiently far away from the boundaries of the interval in which
$\alpha$ is allowed to vary.

One concludes that the mutual information between two patches of the
universe during inflation is of order one, at least in the regimes of
observational relevance. As already mentioned, its dependence on
  the distance $d$ is very mild, especially when
compared to the flat spacetime case, where it is known to decay at its inverse fourth power, see \Refa{Martin:2021xml}. This is because, in the case of inflation (de Sitter
  spacetime), quantum correlations are now produced in Fourier space,
  as revealed by the fact that the quantum state of the perturbations is
  no longer a two-mode coherent state but a two-mode squeezed state,
  which is an entangled state. We will come back to this comparison when we compute the quantum discord later on in this section.
  
  Let us also recall that the mutual
  information between curvature perturbations with opposite Fourier
  modes, $\vec{k}$ and $-\vec{k}$, was computed in
  \Refa{Martin:2015qta} and was found to be of order
\begin{align}
\label{eq:I:k:minus:k}
\mathcal{I}({\vec{k},-\vec{k}})\simeq
4 \log_2\left(H R_k\right)\, ,
\end{align}
where $R_k=a/k$ is the wavelength associated to the mode $k$,
assuming that it is much larger than the Hubble radius (so $H R_k\gg
1)$. For the wavenumbers observed in the CMB, $\ln(H R_k)$ is of order
$50$ at the end of inflation, so $\mathcal{I}({\vec{k},-\vec{k}}) \sim
140$, which is much larger than the typical values encountered in
\Fig{fig:I:infl}. The reason is that, while
$\mathcal{I}({\vec{k},-\vec{k}})$ depends logarithmically on the
relevant scales of the problem, $\mathcal{I}(\vec{x}_1,\vec{x}_2)$
involves the logarithm of the logarithm of those scales, see
\Eq{eq:calI:appr:rough:inflation}. This shows that the amount of
correlations being accessed is smaller in real space than in Fourier
space.

As explained around \Eq{eq:purity:def}, another difference between
correlations in real and Fourier spaces is that, while the curvature
perturbations with opposite wavevectors are placed in a pure state,
and decouple from any other set of opposite wavevectors, in real
space, the system $\{ \vec{x}_1, \vec{x}_2 \}$ is in a mixed state,
since one has implicitly traced over the value of the field at any
other spatial location, to which the system nonetheless couples. This
effective ``self-decoherence'' can be assessed with the purity
parameter $\mathfrak{p}$ defined in \Eq{eq:purity:def}, and the fact
that the symplectic eigenvalues are large in this regime means that
the purity parameter is small, hence that the system we consider here
is strongly mixed. More precisely, the purity associated with the
two-point setup is given by $\mathfrak{p}_{1,2} = 1/(\sigma_+
\sigma_-)\sim (HR)^{-4}$ while the purity for the one-point systems is
given by $\mathfrak{p}_1=\mathfrak{p}_2=1/\sigma_1\sim (HR)^{-2}$.

Finally, although the regime $HR\ll 1$ cannot be probed in the CMB as
argued above, it is still of theoretical interest to discuss this
limit, in order to fully describe the structure of the correlations
present in the field of inflationary perturbations. In this regime, \Eqs{eq:Cov:Inflation:appr}-\eqref{eq:covapprox24} lead to
$\sigma_+\simeq\sigma_-\simeq\sigma_1\simeq
3/(2\pi)\sqrt{1+2\delta-2(1+\delta)\log(\delta/2)}$. Since these three
symplectic values are the same, the mutual information vanishes at
leading order, see \Eq{eq:I:result}. At next-to-leading order, when
$1/\alpha\ll HR\ll 1$, one finds $\mathcal{I} \propto (HR)^4$, and
when $HR\ll 1/\alpha$, one has $\mathcal{I}\propto 1/\alpha^4$, which
coincides with the result obtained in the Minkowski vacuum, see
\Refs{Shiba:2012np, Martin:2021xml}. For comparison, the
mutual information between curvature perturbations with opposite
Fourier modes is given by $\mathcal{I}(\vec{k},-\vec{k})\simeq -
  \log_2(H R_k/2) (H R_k)^2/2$ in this regime $H R_k\ll 1$, see \Refa{Martin:2015qta}, so one
finds that it is suppressed too. Let us note that the fact that the
symplectic eigenvalues are of order one also means that the system $\{
\vec{x}_1 , \vec{x}_2\}$ we consider is almost pure. From
\Eq{eq:purity:def}, one indeed obtains
$\mathfrak{p}_{1,2}=1/(\sigma_+\sigma_-)\simeq 0.56$ and
$\mathfrak{p}_1=\mathfrak{p}_2=1/\sigma_1\simeq 0.75$ with
$\delta=0.1$. This is in contrast with the opposite regime $HR\gg 1$
where we had found that the system is in a strongly mixed state.

\subsubsection{De Sitter quantum discord}
\label{subsubsec:discorddS}

Let us now compute the quantum discord. Having already determined
${\cal I}(\vec{x}_1,\vec{x}_2)$, this means that we need to calculate
${\cal J}(\vec{x}_1,\vec{x}_2)$. In the regime $HR\gg 1$,
\Eqs{eq:sigma:infl:appr}, (\ref{eq:sigmaminf}), (\ref{eq:sigma1inf})
and~(\ref{eq:sigma12inf}) imply that $\sigma_+$, $\sigma_-$ and
$\sigma_1$ are all of order $HR$, while $\sigma_{1\mhyphen 2}$ is of
order $HR/\alpha^2$ and is therefore suppressed compared to
$\sigma_+$, $\sigma_-$ and $\sigma_1$. This means that, in the
quantity appearing below \Eq{eq:E} whose sign
determines which formula one should use for $\mathcal{J}$, and which
is written as the difference of two positive terms, the first term is
of order $(HR)^{10}/\alpha^8$ and the second term is of order
$(HR)^8$. Which term dominates thus depends on how $HR$ compares to
$\alpha^4$, and this implies that the cases $HR\ll\alpha^4$ and $HR\gg
\alpha^4$ need to be distinguished.

Let us first consider the case where $HR\ll\alpha^4$. In this case,
the discriminating quantity appearing in the text after \Eq{eq:E} is
negative, hence the second formula for $E$ needs to be used. This
leads to $E\simeq (\sigma_+\sigma_-/\sigma_1)^2$ at leading order, so
$E$ is of order $(HR)^2$ and is therefore large. Using that
$f(x)\simeq 1/\ln 2+\log_2(x/2)$ when $x\gg 1$, this gives rise to
$\mathcal{J}\simeq \log_2[\sigma_1^2/(\sigma_+ \sigma_-)]$, which
coincides with the expression obtained for $\mathcal{I}$. This means
that, at leading order, there is an exact cancellation between
$\mathcal{I}$ and $\mathcal{J}$. Since both $\mathcal{I}$ and
$\mathcal{J}$ are of order one, this implies that the quantum discord,
which must come from higher-order terms, is a suppressed quantity in
this regime.

In order to compute its value, one needs to work at next-to-leading
order, where $f(x)\simeq 1/\ln 2+\log_2(x/2)-1/(6 x^2 \ln 2)$. For
the mutual information $\mathcal{I}$, \Eq{eq:I:result} yields
\begin{align}
\label{eq:I:nlo:inf}
\mathcal{I}(\vec{x}_1,\vec{x}_2)\simeq
\log_2\left(\frac{\sigma_1^2}{\sigma_+ \sigma_-}\right)
+\frac{1}{6\ln 2}\left(\frac{1}{\sigma_+^2}+\frac{1}{\sigma_-^2}
-\frac{2}{\sigma_1^2}\right)\, ,
\end{align}
while for $\mathcal{J}$, \Eq{eq:J} leads to $E\simeq (\sigma_-
\sigma_+/\sigma_1)^2 +\sigma_{1\mhyphen
  2}^4\sigma_-^2\sigma_+^2/[\sigma_1^2(\sigma_1^4-\sigma_-^2\sigma_+^2)]$,
which gives rise to
\begin{align}
  \mathcal{J}(\vec{x}_1,\vec{x}_2)\simeq
  \log_2\left(\frac{\sigma_1^2}{\sigma_- \sigma_+}\right)
  + \frac{1}{6\ln 2}\left(\frac{\sigma_1^2}{\sigma_-^2 \sigma_+^2}
  -\frac{1}{\sigma_1^2}\right)
  -\frac{1}{2\ln 2}\frac{\sigma_{1\mhyphen 2}^4}
  {\sigma_1^4-\sigma_-^2\sigma_+^2}\, .
\end{align}
In this expression, the second term is of order $(HR)^{-2}$ while the
third term is of order $1/\alpha^8$. Since we work under the
assumption $HR\ll \alpha^4$, the third term can therefore be
neglected, and the quantum discord~\eqref{eq:discord:def} is given by
\begin{align}
\label{eq:discord:appr:inflation:Small:HR}
\mathcal{D}(\vec{x}_1,\vec{x}_2)\simeq \frac{1}{6\ln 2}
\left(\frac{1}{\sigma_+^2}+\frac{1}{\sigma_-^2}-\frac{1}{\sigma_1^2}
-\frac{\sigma_1^2}{\sigma_-^2 \sigma_+^2}\right)\, .
\end{align}
This shows that the discord is of order $(HR)^{-2}$ in this regime. An
explicit expression can be obtained upon
using~\Eq{eq:sigma:infl:appr}, but since it is rather cumbersome, let
us give only its ``rough'' version where the logarithms are assumed to
dominate over terms of order one [\ie similarly to what was done
  in~\Eq{eq:calI:appr:rough:inflation}],
\begin{align} 
\label{eq:discord:appr:inflation:Small:HR:Rough}
\mathcal{D}(\vec{x}_1,\vec{x}_2)\simeq \frac{\pi^2}{12 \ln 2 (HR)^2}
\frac{\ln^2(\alpha \beta)}{ \left\vert \ln\left(\delta/2\right)
  \ln\left(\alpha/2\right)\ln(2\beta)\ln(2\alpha\beta^2)\right\vert}\, .
\end{align}

We now turn to the case $HR\gg\alpha^4$.  In this case, the
discriminating quantity appearing below \Eq{eq:E} is positive, hence the
first formula for $E$ needs to be used. At leading order, it still
gives rise to $E\simeq (\sigma_+\sigma_-/\sigma_1)^2$, so the
cancellation between $\mathcal{I}$ and $\mathcal{J}$ is still
encountered in that case, and the discord is again suppressed. At
next-to-leading order, \Eq{eq:I:nlo:inf} can still be used,
while~\Eq{eq:E} leads to
\begin{align}
\label{eq:Einfcase2}
  E\simeq \left(\frac{\sigma_- \sigma_+}{\sigma_1}\right)^2
  +2\frac{\left\vert \sigma_{1\mhyphen 2}^2\right\vert \sigma_-\sigma_+}{\sigma_1^3}
  +\left(\frac{\sigma_-^2\sigma_+^2}{\sigma_1^4}-1\right)\, .
\end{align}
In this expression, the second term is of order $HR/\alpha^4$ while the third term is of order one.  Since
we work under the assumption that $HR\gg \alpha^4$, the third term can
be discarded, which leads to
\begin{align}
\mathcal{J}(\vec{x}_1,\vec{x}_2)\simeq
\log_2\left(\frac{\sigma_1^2}{\sigma_- \sigma_+}\right)
-\frac{\left\vert\sigma_{1\mhyphen 2}\right\vert^2}
{\sigma_1\sigma_-\sigma_+\ln 2}\, .
\end{align}
The correction to $\mathcal{J}$ is thus of order $1/(HR \alpha^4)$, while the correction to $\mathcal{I}$ is of order $1/(HR)^2$, see
  \Eq{eq:I:nlo:inf}. In the regime $HR\gg \alpha^4$, the correction to
  $\mathcal{J}$ thus provides the dominant contribution, and one
  obtains
\begin{align}
\label{eq:discord:appr:inflation:Large:HR}
\mathcal{D}(\vec{x}_1,\vec{x}_2)
\simeq \frac{\left\vert\sigma_{1\mhyphen 2}^2\right\vert}
        {\sigma_+ \sigma_- \sigma_1\ln 2}. 
\end{align}
The discord is therefore suppressed by $1/(HR \alpha^4)$ in this
regime. More precisely, an explicit expression can be obtained upon
using \Eq{eq:sigma:infl:appr}, and in the ``rough'' limit where the
logarithmic terms dominate over terms of order one, one finds
\begin{align}
\label{eq:discord:appr:inflation:Large:HR:Rough}
\mathcal{D}(\vec{x}_1,\vec{x}_2) \simeq
\frac{2\sqrt{2}\pi}{9\ln 2\, HR \alpha^4}
\frac{\vert \ln(\alpha\beta)\vert}{\vert\ln\left(\delta/2\right)
  \vert^{3/2}\sqrt{\ln\left(\alpha/2\right)\ln(2\beta)\ln(2\alpha\beta^2)}}\, .
\end{align}
\begin{figure}[t]
\begin{center}
\includegraphics[width=0.49\textwidth]{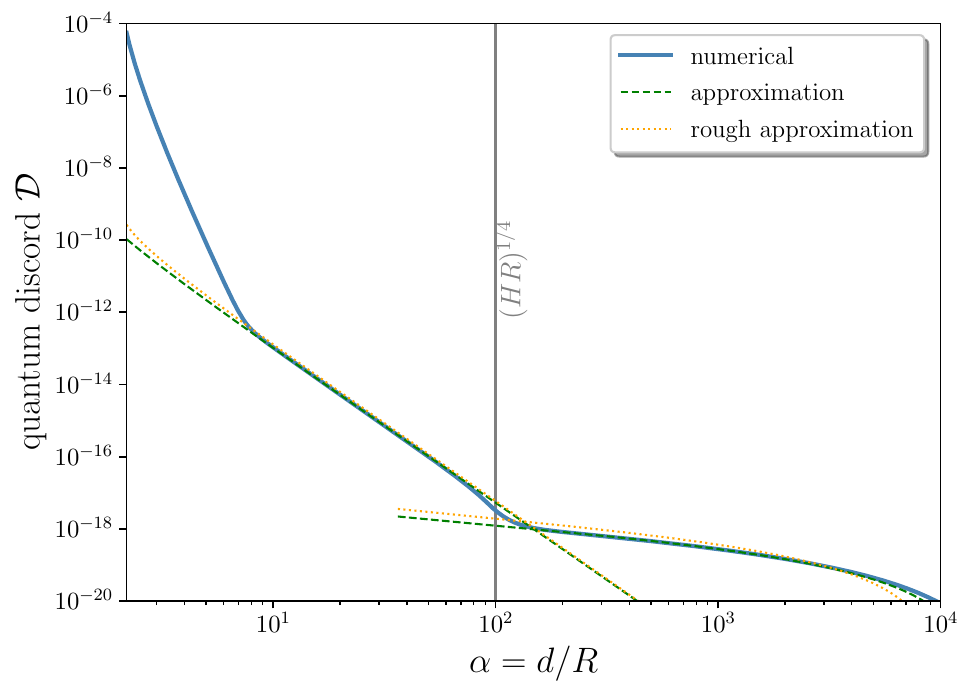}
\includegraphics[width=0.49\textwidth]{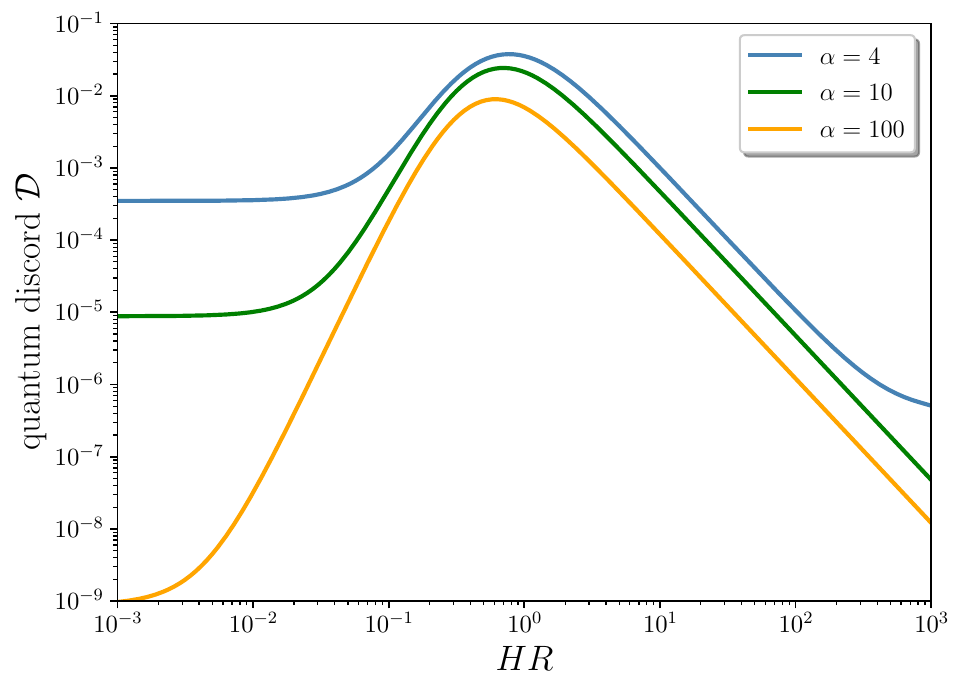}
\caption{Left panel: quantum discord during cosmic inflation for
  $\beta=10^{-4}$, $HR=10^8$ and $\delta=0.1$, as a function of
  $\alpha$, which is varied in the physical range $\alpha\in
  [2(1+\delta),1/\beta]$, corresponding to
  $2R(1+\delta)<d<R_{\mathrm{obs}}$.  The conditions are the same as
  in \Fig{fig:I:infl}. The solid blue curve corresponds to a full
  numerical calculation, the green dashed curves stand for the
  approximated formulas obtained by plugging \Eq{eq:sigma:infl:appr}
  into \Eqs{eq:discord:appr:inflation:Small:HR}
  and~\eqref{eq:discord:appr:inflation:Large:HR} for $HR\ll \alpha^4$
  and $HR\gg \alpha^4$ respectively, while the dotted orange curves
  display the rough
  approximations~\eqref{eq:discord:appr:inflation:Small:HR:Rough}
  and~\eqref{eq:discord:appr:inflation:Large:HR:Rough}. Right panel:
  quantum discord for $\beta=10^{-4}$ and $\delta=0.1$ as a function
  of $HR$, for a few values of $\alpha$ not far from its lower
  bound~\eqref{eq:aplha:cond}.}
\label{fig:discord:infl}
\end{center}
\end{figure}

The above formulas are displayed in the left panel of
\Fig{fig:discord:infl} where they are compared with a numerical
calculation. They are also summarised in \Fig{fig:summary} below. One
can check that, as for the mutual information $\mathcal{I}$, they
provide a good fit to the full result, even when $\alpha$ approaches
its upper bound $\alpha\sim 1/\beta$. When $\alpha$ is close to its
lower bound~\eqref{eq:aplha:cond}, the approximation we have developed
(which assumes $\alpha\gg 1)$ under-estimates the discord, which is
also where the discord reaches its maximal value. For this reason, in
the right panel of \Fig{fig:discord:infl}, we have displayed the
discord for a few values of $\alpha$ not too far from its lower
bound. At large values of $HR$, one recovers the behaviour
$\mathcal{D}\propto 1/(HR)^2$ derived around
\Eq{eq:discord:appr:inflation:Small:HR:Rough}, and at small values of
$HR$, the quantum discord is, like the mutual information
$\mathcal{I}$, of order $1/\alpha^4$ (see the discussion at the
beginning of this section). In between, one can see that it reaches a
maximum value when $HR$ is of order one. This configuration of maximal
discord, where both $\alpha$ and $HR$ are of order one, cannot be
described by our approximations and one needs to resort to a numerical
calculation. 

\subsubsection{Discussion}

At this point, it is worth drawing lessons from the results obtained before. The first conclusion that we reach is that the discord produced in real space during inflation is non-vanishing and that, therefore, ``discord transfer'' does indeed take place from Fourier to real space. 

The second conclusion concerns the amplitude of the discord in real space. Clearly, for reasons already presented, it is small; in any case, smaller than the discord between opposite Fourier modes, which
is half of the mutual information given in \Eq{eq:I:k:minus:k}, see \Refa{Martin:2015qta}, and is therefore large for super-Hubble scales. 

Finally, the real-space mutual information and quantum discord in de Sitter spacetime found above should also be compared to the corresponding result in flat (Minkowski) spacetime, see~\Refa{Martin:2021xml}.
Indeed, to some extent, the Minkowski result represents the benchmark to which other
calculations should be set against. 

In Minkowski Fourier space, no mutual information and no discord is produced because, in flat spacetime, the field always
remains in its vacuum state. As a consequence, in this case, the
mutual information and discord in real space (if present) only originate from the fact that we have
traced out field values at other spatial location. In this sense, the amount of correlations calculated in Minkowski, see~\Refa{Martin:2021xml}, represents the minimum that is always present
in a system due to the passage from Fourier to real space. The
``genuine'' correlations, originating from a physical phenomenon that can
produce entangled quanta in Fourier space (such as interaction
with an exterior classical source, as during inflation), must
therefore appear as an ``additional'' contribution. 

Let us first note that both in Minkowski space and in de-Sitter space, when $\delta\to 0$, the mutual information acquires a finite value (that however depends on the background one considers) while quantum discord vanishes. The reason for this similarity can be understood as follows. When $\delta$ takes a non-vanishing value, the window function in Fourier space decays more rapidly, namely $\widetilde{W}\propto k^{-2}$ if $\delta=0$, see \Eq{eq:Fourier:transform:Heaviside}, while $\widetilde{W}\propto k^{-3}$ if $\delta>0$, see \Eq{eq:Fourier:transform:Heaviside:Improved}. The case $\delta=0$ thus corresponds to larger UV contributions, and since the de-Sitter and Minkowski space-times are identical at small scales, this explains the similar behaviours.

There are however differences. In \Refa{Martin:2021xml}, it was established that, in the Minkowski space-time, both the mutual information and the quantum discord are of order $\alpha^{-4}$ at large distances (unless $\delta=0$, in which case the discord vanishes as mentioned above). In de Sitter, in that limit, we have shown that mutual information remains of order $(HR)^4$ if $HR\ll 1$ and of order one if $HR\gg 1$, and that quantum discord remains of order $(HR)^4$ if $HR\ll 1$ and of order $(HR)^{-2}$ if $HR\gg 1$. Both are therefore larger in de Sitter than in Minkowski.
Therefore, a third conclusion is that, despite the fact that the de Sitter mutual information and discord are smaller in real space than in Fourier space, they are nevertheless always larger than their flat space-time counterpart. This somehow expected result confirms that non-trivial quantum correlations are produced during inflation.  

\subsection{Radiation era}
\label{sec:Radiation}
Let us now study how the correlations contained in the field of
cosmological perturbations evolve after inflation, when the universe
is dominated by a radiation fluid. During that epoch, the scale factor
evolves linearly with conformal time, \ie
$a(\eta)=a_{\mathrm{r}}\left(\eta-\eta_{\mathrm{r}}\right)$. Upon
requiring that the scale factor and its derivative are continuous at
the transition between inflation and the radiation epoch, the two
integration constants $a_{\mathrm{r}}$ and $\eta_{\mathrm{r}}$ can be
determined, and one obtains $\eta_{\mathrm{r}} = 2 \eta_\uend$ and
$a_{\mathrm{r}} = 1/(H_\uend \eta_\uend^2)$, where $\eta_\uend$ and
$H_\uend$ are the values of $\eta$ and $H$ at the end of
inflation. Regarding cosmological perturbations, the Mukhanov-Sasaki
equation~\eqref{eq:Mukhanov:Sasaki} should now be solved with
$\cs=1/3$ and $z= 2\sqrt{3}a\Mp$, and the generic solution reads
\begin{align}
  v_{\vec{ k}}(\eta)= A_{\vec{ k}} \ee^{-ik\frac{\eta-\eta_{\mathrm{r}}}{\sqrt{3}}}
  + B_{\vec{ k}}  \ee^{ik\frac{\eta-\eta_{\mathrm{r}}}{\sqrt{3}}}\, ,
\end{align}
where $A_{\vec{k}}$ and $B_{\vec{k}}$ are two integration constants
that must be set by requiring continuity of the first and second
fundamental forms~\cite{Deruelle:1995kd}. Those matching conditions
take a complicated form in general, but let us recall that the
filtering procedure is such that scales below the coarse-graining
radius, \ie such that $k\gg a/R$, are filtered out. Given that, as
argued below \Eq{eq:Cov:Inflation:appr}, we are very far from
resolving the Hubble radius at the end of inflation, $H_\uend R\gg 1$,
all relevant scales are such that $k\ll a/R\ll a H_\uend$, \ie they
are larger than the Hubble radius at the end of inflation. One can
therefore restrict the analysis of the matching conditions to this
regime, where they simply boil down to requiring the continuity of the
curvature perturbation and of the Bardeen
potential~\cite{Bardeen:1980kt}. At leading order in $k/(a_\uend
H_\uend) = -k \eta_\uend$, this leads to
\begin{align} 
\label{eq:free:mode:rad}
v_{\vec{ k}}(\eta) =
-\frac{3i}{\sqrt{k \epsilon_1}(k\eta_\uend)^2}
\sin\left[\frac{k}{\sqrt{3}}\left(\eta-
    \eta_{\mathrm{r}}\right)\right] ,
\end{align}
where $\epsilon_1$ is the value of the first Hubble-flow parameter
during inflation. The conjugated momentum $p_{\vec{k}} = v_{\vec{k}}'
- (z'/z) v_{\vec{k}}$ is given by
\begin{align}
p_k &= -\frac{3i}{\sqrt{k \epsilon_1}(k\eta_\uend)^2}
\left\lbrace \frac{k}{\sqrt{3}}\cos \left[\frac{k}{\sqrt{3}}\left(\eta-
  \eta_{\mathrm{r}}\right)\right]
- \frac{1}{\eta-\eta_r} \sin\left[\frac{k}{\sqrt{3}}\left(\eta-
    \eta_{\mathrm{r}}\right)\right]  \right\rbrace\, .
\end{align}

\subsubsection{Covariance matrix}
\label{subsubsec:covrad}

Making use of \Eq{PowerSpectrum:interm2}, the reduced power spectra
can then be computed. Upon introducing $u\equiv
k(\eta-\eta_\mathrm{r})/\sqrt{3}$, they are given by
\begin{align}
\calP_{vv}=&\frac{9}{2\pi^2\epsilon_1k^2\eta_\uend^4}\sin^2 u \, , \\
\calP_{vp}=&\frac{3\sqrt{3}}{2\pi^2\epsilon_1k \eta_\uend^4}
\left[\sin u \cos u-\frac{\sin ^2 u}{u}\right]\, , \\
\calP_{pp}=&\frac{3}{2\pi^2\epsilon_1\eta_\uend^4}
     \left[\cos u-\frac{\sin u}{u}\right]^2\, .
\end{align}
One may note that these expressions yield
${\calP}_{vv}{\calP}_{pp}-{\calP}_{vp}^2=0$ while, as argued below
\Eq{eq:PowerSpectra:deSitter}, this combination of the power spectra
should equal $[k^3/(4\pi^2)]^2$. This is because the matching conditions have been
performed at leading order in $k\eta_\uend$ only, while the result
 $[k^3/(4\pi^2)]^2$ comes from higher-order terms. Since they play a negligible role
hereafter, they can be safely neglected.

The reduced power spectra involve power-law and trigonometric
functions of the wavenumber. As a consequence, the entries of the
covariance matrix can still be expressed in terms of the three
integrals $\mathcal{K}(\beta,\mu,\delta)$,
$\mathcal{L}(\beta,\mu,\delta,\rho)$ and
$\mathcal{M}(\beta,\mu,\delta,\rho)$ introduced in \Eq{eq:KLM:def} and
further studied in \App{app:Integrals}. One obtains
 \begin{align}
 \label{eq:gamma11:rad}
 \gamma_{11} = &\frac{6  R^4 H^2 H_\uend^2}{\pi \epsilon_1 G(\delta) }
 \left[  \mathcal{K}\left(\beta,-3,\delta \right)
   -\mathcal{M}\left(\beta,-3,\delta,2A\right)\right],\\
 \gamma_{12} = &\frac{4R^3 H H_\uend^2}{\pi\epsilon_1G(\delta) }
 \left[\mathcal{L}\left(\beta,-1,\delta,2A\right)-\frac{1}{2A^2}
   \mathcal{K}\left(\beta,-3,\delta\right)+\frac{1}{2A^2}\mathcal{M}
   \left(\beta,-3,\delta,2A\right)\right],\\
 \gamma_{22}= & \frac{2R^4 H^2 H_\uend^2}{\pi\epsilon_1  G(\delta)}
 \left[\mathcal{K}\left(\beta,-1,\delta\right)+\frac{1}{A^2}
   \mathcal{K}\left(\beta,-3,\delta\right)+\mathcal{M}
   \left(\beta,-1,\delta,2A\right)
\nonumber  \right.\\ & \left.
-\frac{1}{A^2}\mathcal{M}\left(\beta,-3,\delta,2A\right)
-4\mathcal{L}\left(\beta,-1,\delta,2A\right)\right],
 \\
 \label{eq:gamma13:rad}
 \gamma_{13}= & \frac{6 R^4 H^2 H_\uend^2}{\pi\epsilon_1 G(\delta)}
 \left[\mathcal{L}\left(\beta,-3,\delta,\alpha\right)
   +\frac{2A-\alpha}{2\alpha}\mathcal{L}\left(\beta,-3,\delta,2A-\alpha\right)
  \nonumber\right.\\ & \left.
  -\frac{2A+\alpha}{2\alpha}\mathcal{L}
  \left(\beta,-3,\delta,2A+\alpha\right)\right],
 \\
 \gamma_{14} = & \frac{\sqrt{3}R^4 H^2 H_\uend^2}{\pi\epsilon_1 G(\delta)}
 \left[\frac{1}{\alpha}\mathcal{M}\left(\beta,-3,\delta,2A-\alpha\right)
   -\frac{1}{\alpha}\mathcal{M}\left(\beta,-3,\delta,2A+\alpha\right)
 \nonumber  \right.\\ & \left.
 -\frac{2}{A}\mathcal{L}\left(\beta,-3,\delta,\alpha\right)
 -\frac{2A-\alpha}{A\alpha}\mathcal{L}\left(\beta,-3,\delta,2A-\alpha\right)
 +\frac{2A+\alpha}{A\alpha}\mathcal{L}\left(\beta,-3,\delta,2A+\alpha\right)
 \right],\\
 \label{eq:gamma24:rad}
 \gamma_{24} =& \frac{R^4 H^2 H_\uend^2}{\pi2 \epsilon_1G(\delta) }
 \left[\frac{2}{A\alpha}\mathcal{M}\left(\beta,-3,\delta,2A+\alpha\right)
   -\frac{2}{A\alpha}\mathcal{M}\left(\beta,-3,\delta,2A-\alpha\right) 
  \nonumber    \right.\\ & \left.
 +\frac{2}{A^2}\mathcal{L}\left(\beta,-3,\delta,\alpha\right)
 +2\mathcal{L}\left(\beta,-1,\delta,\alpha\right)
 +\frac{2A-\alpha}{A^2\alpha}\mathcal{L}
 \left(\beta,-3,\delta,2A-\alpha\right)   
   \nonumber    \right.\\ & \left.
-\frac{2A-\alpha}{\alpha}\mathcal{L}\left(\beta,-1,\delta,2A-\alpha\right)   
-\frac{2A+\alpha}{A^2\alpha}
\mathcal{L}\left(\beta,-3,\delta,2A+\alpha\right)   
     \nonumber    \right.\\ & \left.
+\frac{2A+\alpha}{\alpha}\mathcal{L}\left(\beta,-1,\delta,2A+\alpha\right)   
 \right]\, ,
 \end{align}
where we have defined $A =1/({\sqrt{3} H R})$ for notational
convenience. Let us note that contrary to inflation where $H$ is
almost constant, $H$ decreases with time in the radiation era (one has
$H\propto a^{-2}$), so $A$ is a time-dependent parameter.

\subsubsection{Analytical approximations}
\label{subsubsec:approxrad}

The above formulas allow one to compute all relevant quantities
introduced in \Sec{sec:MutualInformation:Discord}. However, as during
inflation, analytical approximations are useful to gain insight in the
result. The considerations presented in \Sec{sec:Inflation} about
$\alpha$, $\beta$ and $\delta$ still apply here, so the regime of
interest is the one where $\beta\ll \alpha\beta\ll 1 \ll
\alpha$. Regarding $HR$, the situation is more subtle. Although, as
argued above, $H_\uend R\gg 1$, $H$ decreases during the radiation
era, hence $HR$ decreases too. At the time of recombination where the
CMB is emitted, $HR$ is large below the first recombination peak, so
for multipoles $\ell <220$, and small above. The two regimes $HR\ll 1$
and $HR\gg 1$ need therefore to be considered. Moreover, in
\Eqs{eq:gamma13:rad}-\eqref{eq:gamma24:rad}, the combinations $2A\pm
\alpha$ appear in the last arguments of some of the $\mathcal{L}$ and
$\mathcal{M}$ integrals, the approximate value of which thus depends
on which of $A\propto (HR)^{-1}$ and $\alpha$ is the largest. This
adds a second pivotal value for $HR$ at $HR=1/\alpha$, so one has to
consider three different regimes depending on the value of $HR$.

Below, we review these three regimes one after the other, making use
of the approximated formulas derived in \App{app:Integrals} for the
integrals $\mathcal{K}$, $\mathcal{L}$ and $\mathcal{M}$, and
employing similar techniques as in \Sec{sec:Inflation}. Let us note
that the approximations of \App{app:Integrals} are valid when the
absolute value of the last argument of the integrals $\mathcal{K}$,
$\mathcal{L}$ and $\mathcal{M}$ is much smaller than $1/\beta$ (since
the expansion in $\beta$ is performed first). As argued in
\Sec{sec:Inflation}, this is the case when the last argument is of
order $\alpha$, and if $A\gg \alpha$, this is also true since $A\beta
\propto (H R_{\mathrm{obs}})\ll 1$ given that observations encompass
many Hubble patches at the time of recombination.

\paragraph{Case $HR\ll 1/\alpha$.}$ $
This regime corresponds
to $HR\ll 1$ and $Hd\ll 1$, so both the size of the patch and the
distance between the two patches lie inside the Hubble
radius. Plugging the approximations of \App{app:Integrals} into
\Eqs{eq:gamma11:rad}-\eqref{eq:gamma24:rad}, one obtains
\begin{align}
\label{eq:gamma:rad:RH:<<1/alpha}
\gamma_{11} \simeq &\frac{2 R^2 H_\uend^2}{\pi \epsilon_1 G(\delta)}
\left[3-2\gamma_\mathrm{E}-2\ln\left(\frac{2\beta}{\sqrt{3} HR}\right)
+\frac{\beta^2}{9H^2R^2}
+\frac{3}{5}(1+\delta)(HR)^2\ln(\sqrt{3}HR)\right],\\
\gamma_{12} \simeq & -\frac{2 R^3 H H_\uend^2}{\pi\epsilon_1 G(\delta)}  
\left[1-\frac{\beta^2}{9H^2R^2}-3\frac{73+53\delta}{200 }H^2R^2
  +3\frac{1+\delta}{5}\ln(\sqrt{3}HR)H^2R^2\right],\\
\gamma_{22}\simeq & \frac{2R^4 H^2 H_\uend^2}{\pi\epsilon_1 G(\delta)}
\left[\frac{3}{4}-\ln\left(\sqrt{3} H R\right) \right],\\
\gamma_{13}\simeq & \frac{2 R^2 H_\uend^2}{\pi\epsilon_1 G(\delta)}
\left[3-2\gamma_\mathrm{E}-2\ln\left(\frac{2\beta}{\sqrt{3} H R}\right)
  +\frac{\beta^2}{9 H^2 R^2}+\frac{\alpha^2}{2}(HR)^2
  \ln\left(\frac{\sqrt{3}}{2}\alpha H R\right)\right],\\
\gamma_{14}\simeq &-\frac{2R^3 H H_\uend^2}{\pi\epsilon_1 G(\delta)} 
\left[1-\frac{\beta^2 }{9 H^2 R^2}-\frac{2\alpha^2}{3}H^2R^2
  +\frac{\alpha^2}{2}H^2R^2 \ln\left(\frac{\sqrt{3}}{2}\alpha H R\right)\right],
 \\
 \gamma_{24}\simeq & \frac{2 R^4 H^2 H_\uend^2}{\pi \epsilon_1G(\delta) }
 \left[ \ln\left(\frac{2}{\sqrt{3} HR \alpha}\right)
   -\frac{1+\delta}{5\alpha^2}-\frac{7\alpha^2}{24}H^2R^2
   -\frac{\alpha^2}{2}H^2R^2
   \ln\left(\frac{2}{\sqrt{3} }\alpha HR\right)\right]\, ,
   \label{eq:gamma:rad:RH:<<1/alpha:end}
\end{align}
where the expansion of $\gamma_{11}$, $\gamma_{12}$, $\gamma_{13}$ and
$\gamma_{14}$ has been performed at next-to-leading order to deal with
the cancellation at leading order when evaluating
$\gamma_{11}-\gamma_{13}$ and $\gamma_{12}-\gamma_{14}$ in the
expression~\eqref{eq:sigma:pm} for $\sigma_-$. These formulas give
rise to the symplectic values
\begin{align}
\label{eq:sigma:rad:SmallHR}
\sigma_+^2\simeq & \left[\frac{4 R^3 H H_\uend^2 }
  {\pi\epsilon_1 G(\delta)}\right]^2
\left\{\left[\frac{3}{2}-\gamma_\mathrm{E}
  -\ln\left(\frac{2\beta}{\sqrt{3} HR}\right)\right]
\left[\frac{3}{4}-\ln\left(\frac{3}{2}\alpha H^2R^2\right)\right]
-1\right\},
\\
\sigma_-^2\simeq & \left[\frac{R^4 H^2 H_\uend^2 \alpha}
  {\pi\epsilon_1 G(\delta)}\right]^22
\ln\left(\frac{2}{\alpha\sqrt{3}HR}\right)
\left[\frac{3}{4}+\ln\left(\frac{\alpha}{2}\right)\right],
 \\
 \sigma_1^2\simeq & \left[\frac{2 R^3 H H_\uend^2}
   {\pi \epsilon_1 G(\delta)}\right]^2
 \left\{\left[3-2\gamma_\mathrm{E}-2
   \ln\left(\frac{2\beta}{\sqrt{3} HR}\right)\right]
 \left[\frac{3}{4}-\ln\left(\sqrt{3} H R\right)\right]-1\right\},
 \\
 \sigma_{1\mhyphen 2}^2\simeq &\left[\frac{2R^3 H H_\uend^2}
   {\pi \epsilon_1 G(\delta)} \right]^2 \left\lbrace
 \ln\left(\frac{2}{\sqrt{3} HR\alpha}\right)
 \left[3-2\gamma_\mathrm{E}-2
   \ln\left(\frac{2\beta}{\sqrt{3} HR}\right)\right]-1 \right\rbrace \, .
\end{align}
In order to estimate the size of these parameters, let us introduce
the length scale $R_\uend$ that corresponds to the Hubble radius at
the end of inflation, so $R_\uend(\eta_\uend)=H_\uend^{-1}$ and
$R_\uend(\eta)=H_\uend^{-1}\, a(\eta)/a_\uend$. As argued below
\Eq{eq:Cov:Inflation:appr}, we are far from resolving such scales, so
\begin{align}
\label{eq:R:gg:Rend}
R\gg R_\uend(\eta)\, .
\end{align}
During the radiation epoch, $H^2(\eta)=H^2_\uend[a_\uend/a(\eta)]^{4}$ and
this allows one to write $\sigma_-\propto R^4 H^2(\eta)
  H_\uend^2\alpha/\epsilon_1 = [R/R_\uend(\eta)]^4
  \alpha/\epsilon_1$. In this expression, $R/R_\uend\gg 1$,
$\alpha>2(1+\delta)$ and $\epsilon_1\ll 1$, therefore $\sigma_-\gg 1$.
The other symplectic values are such that $\sigma_+\sim \sigma_1 \sim
\sigma_{1\mhyphen 2}\sim \sigma_-/(HR \alpha)$. Since we have assume
$HR\ll 1/\alpha$, those parameters are much larger than $\sigma_-$,
hence much larger than one too.

In this regime of large symplectic values, as explained in
\Sec{sec:Inflation}, the one-point and two-point systems are placed in
strongly mixed states. The mutual information can be computed from
\Eq{eq:I:nlo:inf}, which here gives rise to
\begin{align}
  \mathcal{I}\left(\vec{x}_1,\vec{x}_2\right)\simeq \log_2
  \left(\displaystyle\frac{\sqrt{2}}{HR\alpha}\right)
  \displaystyle
  +\frac{1}{2}\log_2\left[
    \frac{\displaystyle\ln\left(\frac{2\beta}{\sqrt{3} HR}\right)
      \ln^2\left(\sqrt{3} HR\right)}{
\ln\left(\displaystyle \frac{3}{2}\alpha H^2R^2\right)
      \ln\left(\displaystyle\frac{2}{\sqrt{3}HR\alpha }\right)
      \ln\left(\displaystyle\frac{\alpha}{2}\right)}\right]\, ,
\end{align}
where we use the same ``rough'' approximation as in
\Sec{sec:Inflation}. A crucial difference with
\Eq{eq:calI:appr:rough:inflation} obtained during inflation is that,
here, the mutual information depends logarithmically on $\alpha HR =
Hd$, and not only on logarithms of logarithms. This means that,
contrary to what happens during inflation, a substantial amount of mutual
information can be accessed during the radiation era.

For the quantum discord, in the regime $\sigma_+\sim\sigma_1\sim
  \sigma_{1\mhyphen 2}\gg \sigma_-\gg 1$, we are in the first
condition of \Eq{eq:J}, which leads to
$E\simeq\sigma_+^2\sigma_-^2/\sigma_1^2$ at leading order, hence
$\mathcal{I}$ and $\mathcal{J}$ cancel out and the discord vanishes at
leading order like during inflation. At next-to-leading order, \Eq{eq:J} yields the same expression as in~\Eq{eq:Einfcase2} for $E$; and although the hierarchy between the symplectic values is different here, it turns out that the dominant term is still the same and we have $E\simeq\sigma_+^2\sigma_-^2/\sigma_1^2+2\sigma_- \sigma_+
\sigma_{1\mhyphen 2}^2/\sigma_1^3$, which gives rise to
$\mathcal{J}\simeq \log_2[\sigma_1^2/(\sigma_+
  \sigma_-)]-\sigma_{1\mhyphen 2}^2/[\sigma_- \sigma_+
  \sigma_1 \ln 2]$. Then, making use of \Eq{eq:I:nlo:inf}, one obtains the same formula as in \Eq{eq:discord:appr:inflation:Large:HR}. In the present context, this formula leads to
\begin{align}
\label{eq:discord:sigmas:LargeHR}
\mathcal{D}\left(\vec{x}_1,\vec{x}_2\right)
\simeq &\frac{\pi G(\delta)}{2\ln 2}
\frac{\epsilon_1}{R^4 H^2 H_\uend^2 \alpha}
\sqrt{\frac{\ln\left(\displaystyle\frac{2}{\sqrt{3}HR\alpha}\right)}
  {\ln\left(\displaystyle\frac{\alpha}{2}\right)
    \ln\left(\displaystyle\frac{3}{2}\alpha H^2R^2\right)
    \ln\left(\sqrt{3}HR\right)}}\, .
\end{align}
One notices that $\mathcal{D}$ is of the same order as $1/\sigma_-$,
and is therefore tiny since we have argued above that $\sigma_-\gg 1$,
see the discussion around \Eq{eq:R:gg:Rend}.

\paragraph{Case $1/\alpha\ll HR\ll 1$}$ $
This regime corresponds
to $HR\ll 1$ but $Hd\gg 1$, so the size of the patches is sub-Hubble
but the distance between the two patches is super-Hubble. In this
regime, the same expressions for $\gamma_{11}$, $\gamma_{12}$ and
$\gamma_{22}$ as those given in \Eqs{eq:gamma:rad:RH:<<1/alpha}-\eqref{eq:gamma:rad:RH:<<1/alpha:end} are
found, and the remaining entries of the covariance matrix are given by
\begin{align}
\label{eq:gamma:RHSmall:but:not:that:much}
\gamma_{13}\simeq &\frac{4 }{\pi\epsilon_1 G(\delta)} R^2 H_\uend^2
\left[1-\gamma_\mathrm{E}-\ln(\alpha\beta)\right] 
\, ,
\\
\gamma_{14}\simeq &  -\frac{4}{9\pi \epsilon_1 G(\delta)}
\frac{R H_\uend^2}{\alpha^2 H} \, ,
\\
\gamma_{24}\simeq & -\frac{8}{81\pi \epsilon_1 G(\delta)\alpha^4 }
\left(\frac{H_\uend}{H}\right)^2\, .
\end{align}
For the symplectic values, the same expression for $\sigma_1$ as the
one given in \Eq{eq:sigma:rad:SmallHR} is obtained, and the remaining
$\sigma_+$, $\sigma_-$ and $\sigma_{1\mhyphen 2}$ parameters read
\begin{align}
  \sigma_+^2\simeq  & \left[\frac{2 R^3 H H_\uend^2 }
    {\pi\epsilon_1 G(\delta)}\right]^2
\left\{\left[5-4\gamma_\mathrm{E}-2\ln\left(\frac{2\alpha\beta^2}
  {\sqrt{3} HR}\right)\right]\left[\frac{3}{4}
  -\ln(\sqrt{3} HR)\right]-1\right\},
\\
\sigma_-^2 \simeq  & \left[\frac{2R^3 H H_\uend^2 }
  {\pi\epsilon_1 G(\delta)}\right]^2
\left\{\left[1+2\ln\left(\frac{\alpha\sqrt{3} HR}{2}\right)\right]
\left[\frac{3}{4}-\ln\left(\sqrt{3}HR\right)\right]-1\right\},
\\
\sigma_{1\mhyphen 2}^2 \simeq & -\left[\frac{4R H_\uend^2}
  {9\pi \epsilon_1 G(\delta) H \alpha^2} \right]^2
\left[3-2\gamma_\mathrm{E}-2\ln\left(\alpha\beta\right)\right]\, . 
\label{eq:simga12:LargeHR}
\end{align}
One thus finds that $\sigma_+$, $\sigma_-$ and $\sigma_1$ are all of
order $R^3 H H_\uend^2/\epsilon_1$ and are therefore large, as argued
below \Eq{eq:R:gg:Rend}. Regarding $\sigma_{1\mhyphen 2}$, it is much
smaller than the other three parameters since one has
$\sigma_{1\mhyphen 2}/\sigma_1\propto (HR\alpha)^{-2}$.
One must therefore expand the mutual information and the
quantum discord in the regime $\sigma_+\sim \sigma_-\sim \sigma_1\gg
\sigma_{1\mhyphen 2}$ and $ \sigma_1\gg
1$.  The one-point and the two-point systems
are still in a strongly mixed state, and making use of \Eq{eq:I:nlo:inf}
for the mutual information, one finds at leading order
\begin{align}
  \mathcal{I}\left(\vec{x}_1,\vec{x}_2\right) \simeq
  \frac{1}{2}\log_2\left[
    \frac{\ln^2\left(\displaystyle
    \frac{\sqrt{3}HR}{2\beta}\right)}
    {\ln\left(\displaystyle\frac{\sqrt{3}HR}{2 \alpha\beta^2}\right)
      \ln\left(\displaystyle\frac{\sqrt{3}}{2} \alpha HR\right)}\right]\, .
\end{align}
As during inflation, the mutual information is given by a logarithm of
a logarithm, and is therefore of order one.

Regarding quantum discord, under the condition~\eqref{eq:R:gg:Rend},
one can show that the discriminating quantity appearing below \Eq{eq:E}
is positive, and that
$E\simeq\sigma_+^2\sigma_-^2/\sigma_1^2+2\sigma_- \sigma_+
\vert\sigma_{1\mhyphen 2}^2\vert/\sigma_1^3$, which gives rise to
$\mathcal{J}\simeq \log_2[\sigma_1^2/(\sigma_+
  \sigma_-)]-\vert\sigma_{1\mhyphen 2}^2\vert/[\sigma_- \sigma_+
  \sigma_1\ln 2]$. The same expression of $\mathcal{D}$ in terms of the
symplectic values is obtained as the one given in
\Eq{eq:discord:sigmas:LargeHR}, which here reduces to
\begin{align}
\mathcal{D}\left(\vec{x}_1,\vec{x}_2\right)
\simeq&\frac{\sqrt{2}\pi G(\delta)}{81\ln 2}
\frac{\epsilon_1}{R^7 H_\uend^2 H^5 \alpha^4}
\frac{\left\vert \ln\left(\alpha\beta\right) \right\vert
  \left\vert \ln\left(\sqrt{3}HR\right)\right\vert^{-3/2}}
     {\sqrt{\ln\left(\displaystyle\frac{\sqrt{3} HR}{2\alpha \beta^2}\right)
         \ln\left(\displaystyle\frac{\sqrt{3}}{2}\alpha HR\right)
         \ln\left(\displaystyle\frac{\sqrt{3}HR}{2\beta}\right)}}\, .
\end{align}
One thus finds that the discord is of order $1/(\sigma_1
R^4H^4\alpha^4)$ in this regime. Since both $\sigma_1$ and $HR\alpha$
are large, this means that the discord is again tiny.
\paragraph{Case $HR\gg 1$}$ $
In this case, both the size
of the patches and the distance between them is larger than the Hubble
radius. Plugging the approximations of \App{app:Integrals} into
\Eqs{eq:gamma11:rad}-\eqref{eq:gamma24:rad}, one obtains
\begin{align}
  \gamma_{11} \simeq &\frac{R^2 H_\uend^2 }{\pi \epsilon_1 G(\delta)}
  \left[7-4\gamma_\mathrm{E}-4\ln(2\beta )-2\delta\right]\, , \\
  \gamma_{12} \simeq & -\frac{1-\delta}{\pi  \epsilon_1 G(\delta)}
  \frac{R  H_\uend^2}{H}, \\
  \gamma_{22}\simeq & \frac{1}{9\pi\epsilon_1 G(\delta)}
  \left(\frac{H_\uend}{H}\right)^2
  \left[1-\delta-2(1-2\delta)
    \log \frac{\delta}{2}\right]\, ,
\end{align}
while $\gamma_{13}$, $\gamma_{14}$ and $\gamma_{24}$ are still given
by \Eq{eq:gamma:RHSmall:but:not:that:much}. From these expressions,
the symplectic values can be approximated by
\begin{align}
  \sigma_+^2\simeq &\left[ \frac{R H_\uend^2}
    {3\pi\epsilon_1 G(\delta)H}\right]^2
  \biggl\{\left[11-8\gamma_\mathrm{E}-4\ln(2\alpha\beta^2)-2\delta\right]
  \left[1-\delta-2(1-2\delta)\ln \frac{\delta}{2}\right]
 \nonumber \\ &
  -9\left(1-2\delta\right)\biggr\},
  \\
  \sigma_-^2 \simeq & \left[\frac{R H_\uend^2}
    {3\pi\epsilon_1 G(\delta)H}\right]^2
 \left\{\left(3+4\ln \frac{\alpha}{2}-2\delta\right)
 \left[1-\delta-2(1-2\delta)\ln \frac{\delta}{2}\right]
 -9\left(1-2\delta\right)\right\},
 \\
 \sigma_1^2\simeq &\left[\frac{R H_\uend^2}
   {3\pi\epsilon_1 G(\delta) H}\right]^2
 \biggl\{\left[7-4\gamma_\mathrm{E}-4\ln(2\beta )-2\delta\right]
 \left[1-\delta-2(1-2\delta)\log \frac{\delta}{2}\right]
 -9(1-2\delta)\biggr\}\, ,
\end{align}
where $\sigma_{1\mhyphen 2}$ is still given by
\Eq{eq:simga12:LargeHR}.  We are therefore in the regime where
$\sigma_1\sim\sigma_+\sim \sigma_-\gg \sigma_{1\mhyphen 2}\gg 1$. The
situation is thus the same as in the previous case, namely the case
$1/\alpha\ll HR\ll 1$. The mutual information is thus given by
\Eq{eq:I:nlo:inf}, which here leads to
\begin{align}
\mathcal{I}(\vec{x}_1,\vec{x}_2)\simeq 
\frac{1}{2}\log_2\left[\frac{\ln^2\left(2\beta\right)}
  {-\ln\left(2\alpha\beta^2\right)
    \ln\left(\alpha/2\right)}\right]
\end{align}
at leading order, and the quantum discord can be obtained from
\Eq{eq:discord:sigmas:LargeHR}, which reduces to
\begin{align}
\mathcal{D}\left(\vec{x}_1,\vec{x}_2\right) 
\simeq&\frac{\sqrt{2}\pi G(\delta)}{3\ln 2}
\frac{H\epsilon_1}{R H_\uend^2 \alpha^4}
\frac{\left\vert \ln(\alpha\beta)\right\vert \left \vert
  \ln\left(\delta/2\right) \right\vert^{-3/2}}
     {\sqrt{\ln\left(2\alpha\beta^2\right)
         \ln\left(\alpha/2\right)\ln\left(2\beta\right)}}\, .
\end{align}
One thus finds that the mutual information is of order one, while
quantum discord is strongly suppressed in this regime.
\begin{figure}[t]
\begin{center}
\includegraphics[width=0.485\textwidth]{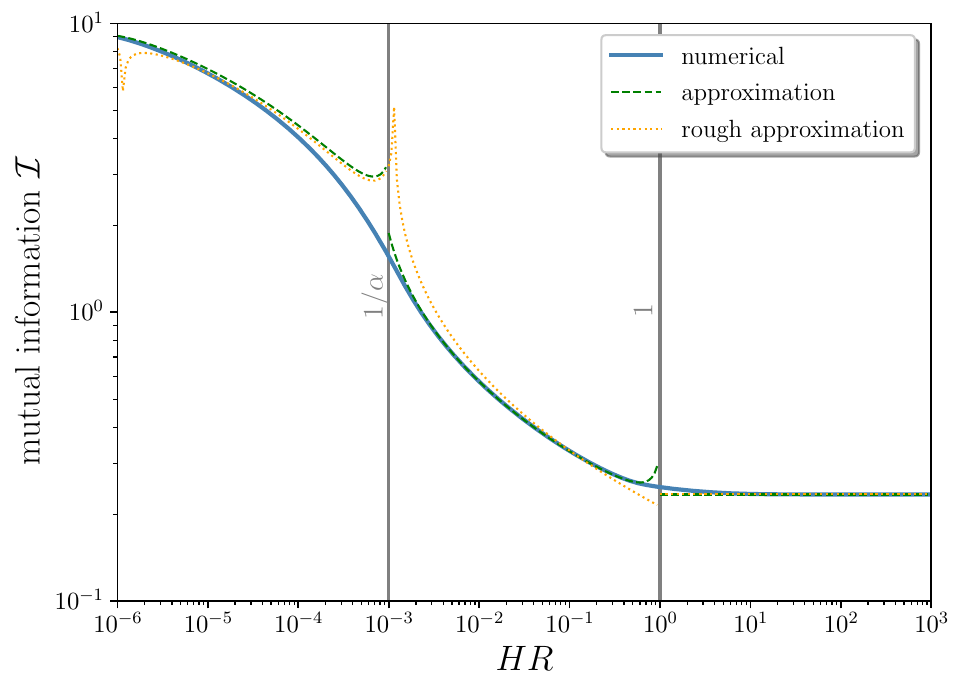}
\includegraphics[width=0.495\textwidth]{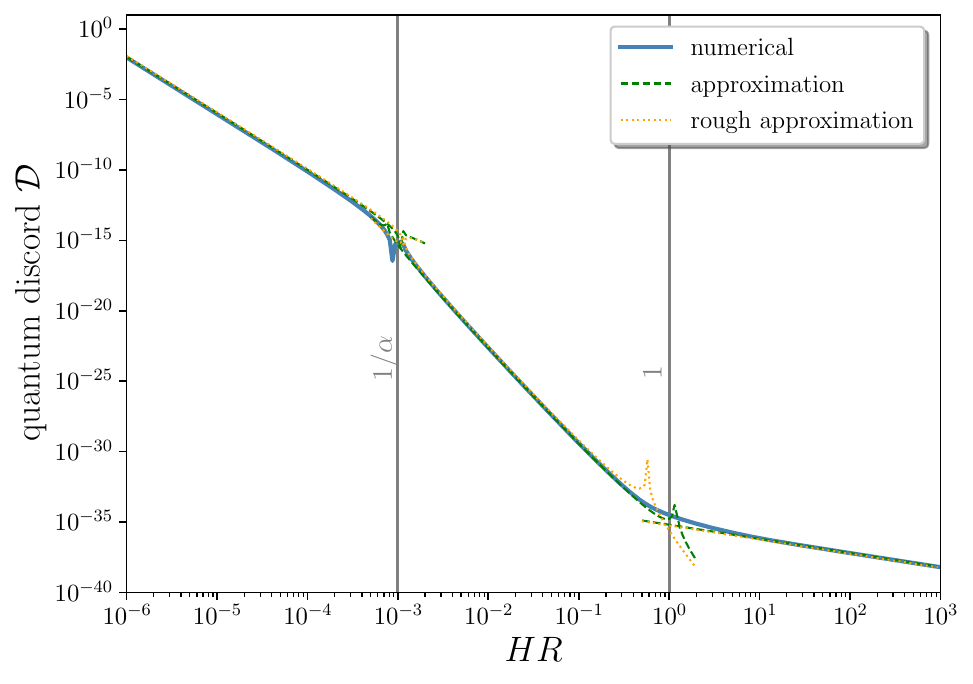}
\caption{Mutual information $\mathcal{I}(\vec{x}_1,\vec{x}_2)$ (left
  panel) and quantum discord $\mathcal{D}(\vec{x}_1,\vec{x}_2)$ (right
  panel) during the radiation era as a function of the coarse-graining
  radius $R$, in units of the Hubble radius, for $\alpha=10^3$,
  $\beta=10^{-6}$, $\delta=0.1$, $\epsilon_1=10^{-2}$ and
  $H/H_\uend=10^{-10}$. The first regime, $HR\ll 1/\alpha$,
  corresponds to the situation where both the experimental resolution
  $R$ and the distance $d$ between the two measured patches are within
  the Hubble distance. The second regime, $1/\alpha\ll RH\ll 1$,
  corresponds to sub-Hubble patches distant by more than the Hubble
  radius while in the third regime, $RH\gg 1$, both the size of the
  patches and their distance is larger than the Hubble radius. The
  lower bound $HR>\beta$ comes from the condition
  $R_{\mathrm{obs}}>H^{-1}$, \ie we observe more than a Hubble patch
  at the time of recombination.}
\label{fig:radiation}
\end{center}
\end{figure}

The above analytical approximations are compared with a numerical
evaluation of the full formulas in \Fig{fig:radiation}. One can check
that they provide indeed a good description of the full result. This
confirms that, at the scales observed in the CMB, the 
mutual information and/or the quantum discord in real
space are non-vanishing but suppressed compared to their typical values in Fourier space. The formulas derived in this section are also summarised in
\Fig{fig:summary} where order-one and logarithmic prefactors have been
removed for clarity.

\section{Conclusion}
\label{sec:conclusions}
\begin{figure}[t]
\begin{center}
\includegraphics[width=0.80\textwidth]{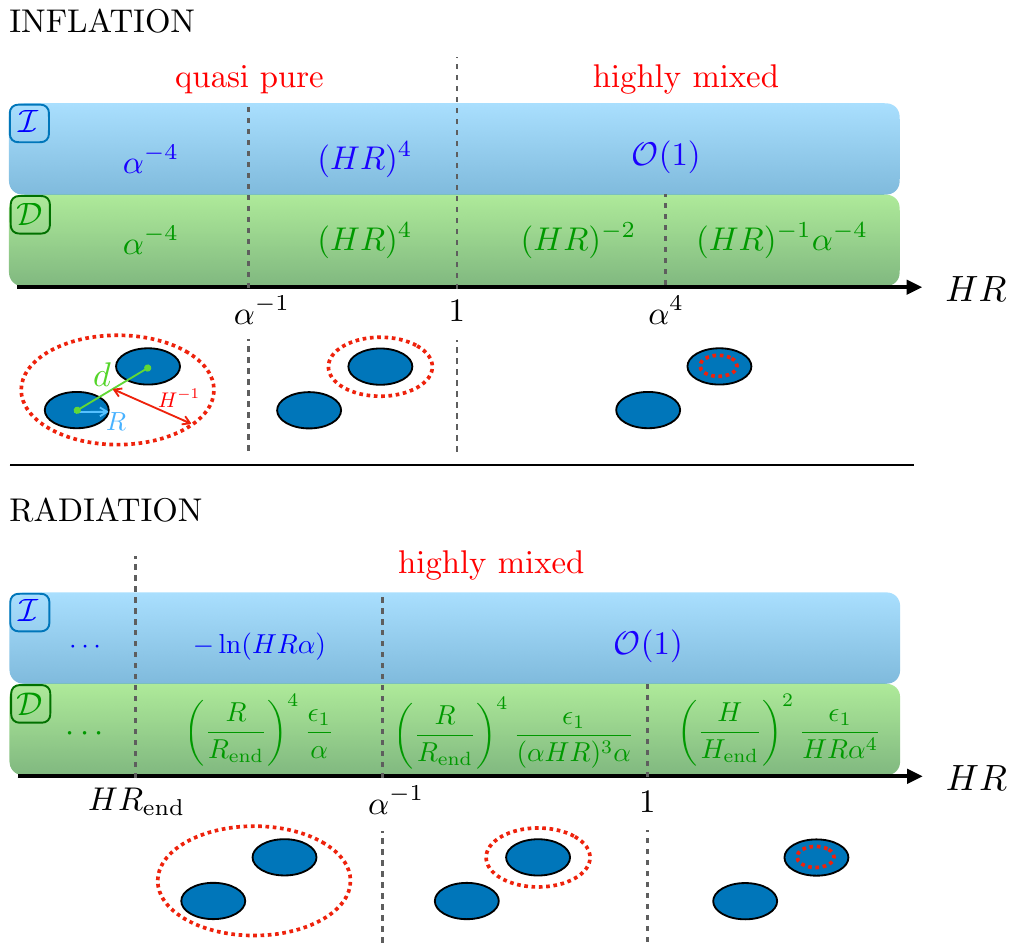}
\caption{Summary of the main results found for the mutual information $\mathcal{I}(\vec{x}_1,\vec{x}_2)$ and the quantum discord $\mathcal{D}(\vec{x}_1,\vec{x}_2)$ contained in the field of cosmological perturbations during the inflation (top panel) and radiation-dominated (bottom panel) eras. The results are given without order-one and logarithmic prefactors. We recall that $R$ is the size of the patch over which the field is measured (it can be thought of as the spatial resolution of the experiment), $H$ is the Hubble parameter and $H_\uend$ its value at the end of inflation, $\alpha$ is the ratio between the distance separating the two measured patches at $\vec{x}_1$ and $\vec{x}_2$ and $R$, $\epsilon_1$ is the value of the first Hubble-flow parameter during inflation, and $R_\uend$ corresponds to the Hubble radius at the end of inflation and properly redshifted at any other time. When $HR\ll \alpha^{-1}$ (respectively $HR\gg \alpha^{-1}$), the distance separating the two measured patches is smaller (respectively larger) than the Hubble radius.}
\label{fig:summary}
\end{center}
\end{figure}
In this work, we have calculated, in real space, the mutual information $\mathcal{I}$ and the quantum discord $\mathcal{D}$ of cosmological curvature perturbations, during an early phase of cosmic inflation and during the subsequent radiation era. Our goal was to give a first estimate of the amount of quantum correlations present in the CMB, in real space. In order to carry out this task, we have used the framework outlined in Ref.~\cite{Martin:2021xml}, which provides a mean to calculate, in real space, the mutual information $\mathcal{I}$ and the quantum discord $\mathcal{D}$ contained in free quantum fields.
We derived explicit and exact analytical expressions, for which we then obtained analytical approximations, confirmed by numerical computations. 

Our main results are summarised in \Fig{fig:summary}. During inflation, when the distance $d=\vert\vec{x}_1-\vec{x}_2\vert$ between the two measured patches of size $R$ is smaller than the Hubble radius $H^{-1}$, both $\mathcal{I}$ and $\mathcal{D}$ are non-vanishing but suppressed by $(R/d)^4$. This coincides with the result obtained in the Minkowski vacuum, see \Refa{Shiba:2012np} and \Refa{Martin:2021xml}. When $d$ is larger than the Hubble radius but $R$ is not, both $\mathcal{I}$ and $\mathcal{D}$ are suppressed by $(R/H^{-1})^4$. Otherwise, when both the size of the patches and the distance between them is larger than the Hubble radius, the mutual information is of order one while quantum discord is suppressed by inverse powers of $(R/H^{-1})^4$ and $d/R$. During the radiation epoch, the mutual information $\mathcal{I}$ can be substantial when $d<H^{-1}$, and is of order one otherwise. However, the quantum discord is always highly suppressed, at least when $R$ is larger than the redshifted size of the Hubble radius at the end of inflation (recall that we are far from probing such scales). Note that in the specific case where the window function is sharp in real space ($\delta=0$), both in Minkowski and in de Sitter, we found that the quantum discord vanishes.

Our main conclusion is therefore that, with measurements of the CMB, for reasonable (current and future) spatial resolution, even though a non-vanishing amount of entanglement entropy or quantum discord in real space exists, these
quantities are highly suppressed.

This statement may seem at odd with the fact that in Fourier space, free quantum fields evolving on curved backgrounds undergo creation of pairs of entangled particles with wavevectors $\vec{k}$ and $-\vec{k}$, such that for the bipartite system $\{ \vec{k}, -\vec{k} \}$, a very substantial amount of mutual information and quantum discord is reached on super-Hubble scales~\cite{Martin:2015qta}. A crucial difference between the $\{ \vec{k}, -\vec{k} \}$ and the $\{ \vec{x}_1, \vec{x}_2 \}$ systems is however that, since different Fourier modes decouple for a free field evolving on a homogeneous background, the systems $\{ \vec{k}, -\vec{k} \}$ are placed in a pure state, and the quantum state of the full field is a direct product of pure states, one for each set $\{ \vec{k}, -\vec{k} \}$. In real space however, correlations build up between the field configuration at different spatial locations. As a consequence, by considering the system $\{ \vec{x}_1, \vec{x}_2 \}$, one implicitly traces over the configuration of the field at any location different from $\vec{x}_1$ and $\vec{x}_2$, which implies that the bipartite system $\{ \vec{x}_1, \vec{x}_2 \}$ is placed in a mixed state. This effective ``self-decoherence'' leads to a suppression of quantum discord (at the technical level, we have seen that large symplectic eigenvalues are indeed both associated with a small purity parameter and to an exact cancellation between $\mathcal{I}$ and $\mathcal{J}$ at leading order). 

The situation depicted in \Fig{fig:summary} could therefore be summarised as follows. For scales remaining inside the Hubble radius throughout the entire cosmic evolution, there is no creation of pairs of entangled particles, hence, although the real-space bipartite system is in a quasi pure state, the mutual information and the quantum discord remain small. For scales stretched above the Hubble radius, a large amount of entangled particles are created in each Fourier mode, but ``self-decoherence'' leads to an important suppression of the quantum discord. 

There is, however, one exception to this general conclusion, namely those scales that cross the Hubble radius at the end of inflation, and that undergo particle creation only for a brief period around that time. The real-space bipartite systems at those scales are in a quasi pure state, and if $d$ is not much larger than $R$ (namely the two measured patches are close one to another, in units of their size), both the mutual information and the quantum discord is of order one or slightly below, see the right panel of \Fig{fig:discord:infl}. Such scales are too small to be accessed in CMB measurements, as well as in measurements of the large-scale structures performed at smaller redshift. The only possibility would be that primordial black holes~\cite{Carr:1974nx} form straight after the end of inflation from the amplification of those scales, as may occur \eg as a result of the preheating instability~\cite{Jedamzik:2010dq, Martin:2019nuw, Martin:2020fgl, Auclair:2020csm}. 

The conclusions reached in the present article should also be compared to what is obtained in Minkowski (flat) spacetime. In this case, there is no quantum discord at all in Fourier space since 
the fields always remain in their vacuum state. In real space, the discord also vanishes for sharp window functions but can be non-zero for smoother window functions, in which case it scales as the inverse of the fourth power of the distance between the two patches~\cite{Martin:2021xml}. Summarising, the mutual information and quantum discord found in cosmology are, unsurprisingly, always larger than in flat spacetime.

It may also be noticed that estimating or gauging how much of mutual information and/or of quantum discord is needed to be able to measure quantum correlations in a system is not obvious (even if, of course, the larger the discord, the higher the chance to observe a signature) since it may depend on the system, the experimental protocol, the precision of the detector and so on. Therefore, a priori, a small amount of quantum discord does not necessarily rule out the possibility to highlight the quantum origin of the perturbations in real space. On the other hand, the fact that the discord is not vanishing does not guarantee, even in principle, that quantum effects can be detected. For instance, it remains to be seen if the Bell's inequality can be violated in real space, since a non-vanishing discord for mixed states does not necessarily imply quantum entanglement~\cite{2010PhRvL.105b0503G}.

To close this article, let us acknowledge that the results obtained 
here certainly indicate that revealing the quantum origin of the cosmological perturbations is not an easy task, especially at CMB scales. We have argued that it might be more feasible on much smaller scales. However, even if ultra-light black holes were detected (possibly through the emission of an associated gravitational-waves background~\cite{Papanikolaou:2020qtd}), it would remain to determine which measurable quantities could unveil the quantum nature of the underlying overdensity field, and we leave this discussion to future work.
 

\appendix

\section{Approximation for the trigonometric integrals}
\label{app:Integrals}

In this appendix, we explain how to calculate and approximate the
three integrals~(\ref{eq:KLM:def}), (\ref{eq:defL})
and~(\ref{eq:defM}) defining the functions ${\cal
  K}(\beta,\mu,\delta)$, ${\cal L}(\beta,\mu,\delta,\rho)$ and ${\cal
  M}(\beta,\mu,\delta,\rho)$. Since the window
function~\eqref{eq:Fourier:transform:Heaviside:Improved} involves
trigonometric and power-law functions of $k$, these three integrals
can be expressed solely in terms of 
\begin{align}
\label{eq:CS:def}
C_\gamma(\nu,\beta) &\equiv    \vert\nu\vert^{\gamma-1}
\int _{\vert \nu \vert \beta}^{+\infty}\frac{\dd u}{u^\gamma}\cos u,\\
S_\gamma(\nu,\beta)&\equiv \mathrm{sign}(\nu)
\vert\nu\vert^{\gamma-1}\int_{\vert \nu \vert \beta}^{+\infty}
\frac{\dd u}{u^\gamma}\sin u.
\end{align}
Explicitly, after long but straightforward calculations, one indeed
arrives at the following expressions
\begin{align}
\label{eq:calK:CS}
 {\cal K}(\beta,\mu,\delta)& = 
\frac{9}{\delta^2{\cal F}^2(\delta)}\Bigg[
\left(1+\delta+\frac{\delta^2}{2}\right)
\frac{\beta^{\mu-5}}{5-\mu}
+4\frac{\beta^{\mu-7}}{7-\mu}
-\frac{1}{2}C_{6-\mu}(2,\beta)
+2C_{8-\mu}(2,\beta)
\nonumber \\ &
+2 S_{7-\mu}(2,\beta)
-(1+\delta)
C_{6-\mu}(\delta, \beta)
-4C_{8-\mu}(\delta, \beta)
-2\delta S_{7-\mu}(\delta, \beta)
\nonumber \\ &
+2(1+\delta)
S_{7-\mu}\left(2+2\delta,\beta\right)
-\frac{\left(1+\delta\right)^2}{2}
C_{6-\mu}\left(2+2\delta,\beta\right)
-4C_{8-\mu}\left(2+\delta,\beta\right)
\nonumber \\ &
+2
C_{8-\mu}\left(2+2\delta,\beta\right)
+(1+\delta)C_{6-\mu}\left(2+\delta,\beta\right)
-2(2+\delta)S_{7-\mu}\left(2+\delta,\beta\right)
\Bigg],\\
{\cal L}(\beta,\mu,\delta,\rho)& = 
\frac{9}{\rho \delta^2{\cal F}^2(\delta)}\Bigg\{
\left(1+\delta+\frac{\delta^2}{2}\right)
S_{7-\mu}(\rho, \beta)
+4 S_{9-\mu}(\rho, \beta)
\nonumber \\ & 
-\frac{1}{4}S_{7-\mu}\left(\rho+2,\beta\right)
-\frac{1}{4}S_{7-\mu}\left(\rho-2,\beta\right)
+S_{9-\mu}\left(\rho+2,\beta\right)
\nonumber \\ & 
+S_{9-\mu}\left(\rho-2,\beta\right)
+C_{8-\mu}\left(\rho-2,\beta\right)
-C_{8-\mu}\left(\rho+2,\beta\right)
\nonumber \\ &
-\frac{1+\delta}{2}
\left[S_{7-\mu}\left(\delta+\rho,\beta\right)
+S_{7-\mu}\left(\rho-\delta,\beta\right)\right]
-2
S_{9-\mu}\left(\delta+\rho,\beta\right)
\nonumber \\ & 
-2S_{9-\mu}\left(\rho-\delta,\beta\right)
-\delta
C_{8-\mu}\left(\rho-\delta,\beta\right)
+\delta C_{8-\mu}\left(\delta+\rho,\beta\right)
\nonumber \\ & 
+(1+\delta)
\left[C_{8-\mu}\left(\rho-2-2\delta,\beta\right)
-C_{8-\mu}\left(2+2\delta+\rho,\beta\right)\right]
\nonumber \\ & 
-\frac{(1+\delta)^2}{4} 
\left[S_{7-\mu}\left(2+2\delta+\rho,\beta\right)
+S_{7-\mu}\left(\rho-2-2\delta,\beta\right)\right]
\nonumber \\ & 
+ S_{9-\mu}\left(2+2\delta+\rho,\beta\right)
+S_{9-\mu}\left(\rho-2-2\delta,\beta\right)
\nonumber \\ & 
+\frac{1+\delta}{2}
\left[S_{7-\mu}\left(2+\delta+\rho,\beta\right)
+S_{7-\mu}\left(\rho-2-\delta,\beta\right)\right]
\nonumber \\ &
-2
S_{9-\mu}\left(2+\delta+\rho,\beta\right)
-2S_{9-\mu}\left(\rho-2-\delta,\beta\right)
\nonumber \\ & 
-(2+\delta)
\left[C_{8-\mu}\left(\rho-2-\delta,\beta\right)
-C_{8-\mu}\left(2+\delta+\rho,\beta\right)\right]\Bigg\},\\
\mathcal{M}(\beta,\mu,\delta,\rho) & =  \frac{1}{36\delta^2\mathcal{F}^2(\delta)}\Big[
2 \left(\delta ^2+2 \delta +2\right)C_{6-\mu}\left(\rho,  \beta\right)
+2 (\delta +1)C_{6-\mu}\left(\rho -\delta -2, \beta\right)
\nonumber \\ & 
-2 (\delta +1) C_{6-\mu}\left( \rho -\delta , \beta\right)
-2 (\delta +1) C_{6-\mu}\left(\rho +\delta ,\beta \right)
\nonumber \\ & 
-(\delta +1)^2  C_{6-\mu}\left(\rho +2 \delta +2, \beta\right)
-(\delta +1)^2 C_{6-\mu}\left(\rho -2 \delta -2 ,\beta \right)
\nonumber \\ & 
-8 C_{8-\mu}\left(\rho -\delta -2,\beta\right)
-8C_{8-\mu}\left(\rho -\delta ,\beta\right)
-8 C_{8-\mu}\left(\rho +\delta ,\beta\right)
\nonumber \\ & 
-8 C_{8-\mu}\left(\rho +\delta +2,\beta\right)
+4 C_{8-\mu}\left(\rho +2 \delta +2,\beta\right)
- C_{6-\mu}\left(\rho -2, \beta\right)
\nonumber \\ & 
- C_{6-\mu}\left(\rho +2, \beta\right)
+4 C_{8-\mu}\left(\rho -2, \beta\right)
+16 C_{8-\mu} \left(\rho,  \beta\right)
\nonumber \\ & 
+4 C_{8-\mu}\left(\rho +2, \beta\right)
+2 (\delta +1) C_{6-\mu}\left(\rho +\delta +2,\beta\right)
\nonumber \\ & 
+4 C_{8-\mu}\left(\rho -2 \delta -2,\beta\right)
+4 (\delta +2) S_{7-\mu}\left(\rho -\delta -2, \beta\right)
\nonumber \\ & 
+4 \delta S_{7-\mu}\left(\rho -\delta ,\beta\right)
-4 \delta S_{7-\mu}\left(\rho +\delta ,\beta\right)
-4 (\delta +2)S_{7-\mu}\left(\rho +\delta +2,\beta\right)
\nonumber \\ & 
+4 (\delta +1)S_{7-\mu}\left(\rho +2 \delta +2,\beta\right)
-4 (\delta +1) S_{7-\mu}\left(\rho -2 \delta -2,\beta\right)
\nonumber \\ & 
-4S_{7-\mu}\left(\rho -2, \beta\right)
+4S_{7-\mu}\left(\rho +2, \beta\right)
\Big] .
\end{align}

The functions $C_\gamma(\nu,\beta)$ and $S_\gamma(\nu,\beta)$ can be
performed in terms of the cosine integral function,
\begin{align}
  \mathrm{Ci}(z) = -\int_z^\infty \frac{\cos(t)}{t}\dd t\, ,
\end{align}
when $\gamma$ is a positive integer number. In practice, the values of
$\gamma$ that are relevant for the calculation presented in the main
text are $\gamma=3,\, 5,\, 7,\, 9,\, 11$ for $C_\gamma(\nu,\beta)$,
and $\gamma=4,\, 6,\, 8,\, 10,\, 12$ for $S_\gamma(\nu,\beta)$. For
those values, one has
\begin{align}
C_3(\nu,\beta) =& 
\frac{\nu^2}{2} \text{Ci}( \vert\nu\vert \beta)-\frac{\nu}{2\beta}
\sin ( \nu \beta)+\frac{1}{2\beta^2}\cos (\nu \beta),
 \\
C_5(\nu,\beta) =& 
-\frac{\nu^4}{24} \mathrm{Ci}(\vert\nu\vert \beta)
+\frac{\nu^2\beta^2-2}{24 \beta^4} \nu \beta \sin (\nu \beta)
+\frac{6-\nu^2\beta^2}{24\beta^4} \cos (\nu \beta),
\\
C_7(\nu,\beta) =& 
\frac{\nu^6}{720} \mathrm{Ci}(\vert \nu\vert \beta)
-\frac{\nu^4\beta^4-2 \nu^2\beta^2+24}{720 \beta^6}\nu \beta \sin (\nu \beta)
\nonumber \\ & 
+\frac{\nu^4\beta^4-6 \nu^2\beta^2+120}{720 \beta^6} \cos (\nu \beta),
\\
C_9(\nu,\beta) =& 
-\frac{\nu^8}{40320 } \mathrm{Ci}(\vert\nu\vert \beta)
+\frac{\nu^6 \beta^6-2 \nu^4 \beta^4
  +24 \nu^2 \beta^2-720}{40320 \beta^8}\nu  \beta \sin (\nu \beta)
\nonumber \\ & 
-\frac{\nu^6 \beta^6-6 \nu^4 \beta^4
  +120 \nu^2 \beta^2-5040}{40320 \beta^8} \cos (\nu \beta),
\\
C_{11}(\nu,\beta) =& 
\frac{\nu^{10}}{3628800} \mathrm{Ci}(\vert\nu\vert \beta)
\nonumber \\ & 
-\frac{\nu^8\beta^8-2 \nu^6\beta^6
  +24 \nu^4\beta^4-720 \nu^2\beta^2+40320}{3628800 \beta^{10}}
\nu \beta \sin (\nu \beta)
\nonumber \\ & 
+\frac{\nu^8\beta^8-6 \nu^6\beta^6+120 \nu^4\beta^4
-5040 \nu^2\beta^2+362880}{3628800 \beta^{10}} \cos (\nu \beta),
\\
S_4(\nu,\beta) =& 
\frac{\nu^3 }{6 } \mathrm{Ci}( \vert\nu\vert \beta)
-\frac{\nu^2 \beta^2-2}{6  \beta^3} \sin (\nu \beta)
+\frac{\nu \beta}{6  \beta^3} \cos (\nu \beta),
\\
S_6(\nu,\beta) =& 
-\frac{\nu^5}{120 } \mathrm{Ci}(\vert\nu\vert \beta)
+\frac{\nu^4\beta^4-2 \nu^2\beta^2+24}{120 \beta^5} \sin (\nu \beta)
\nonumber \\ & 
+\frac{6-\nu^2\beta^2}{120 \beta^5} \nu \beta \cos (\nu \beta),
\\
S_8(\nu,\beta) =& 
\frac{\nu^7 }{5040} \mathrm{Ci}(\vert\nu\vert \beta)
-\frac{\nu^6\beta^6-2 \nu^4\beta^4+24 \nu^2 \beta^2-720}{5040 \beta^7}
\sin (\nu \beta)
\nonumber \\ & 
+\frac{\nu^4 \beta^4-6 \nu^2 \beta^2+120}{5040 \beta^7}
\nu \beta \cos (\nu \beta),
\\
S_{10}(\nu,\beta) =& 
-\frac{\nu^9}{362880} \mathrm{Ci}(\vert\nu\vert \beta)
\nonumber \\ & 
+\frac{\nu^8 \beta^8-2 \nu^6 \beta^6+24 \nu^4 \beta^4
  -720 \nu^2 \beta^2+40320}{362880 \beta^9} \sin (\nu \beta)
\nonumber \\ & 
-\frac{\nu^6 \beta^6-6 \nu^4 \beta^4
  +120 \nu^2 \beta^2-5040}{362880 \beta^9}\nu \beta \cos (\nu \beta),
\\
S_{12}(\nu,\beta) =& 
\frac{\nu^{11}}{39916800} \mathrm{Ci}(\vert\nu\vert \beta)
\nonumber \\ & 
-\frac{\nu^{10}\beta^{10}-2 \nu^8 \beta^8+24 \nu^6 \beta^6
  -720 \nu^4 \beta^4+40320 \nu^2 \beta^2-3628800}{39916800 \beta^{11}}
\sin (\nu \beta)
\nonumber \\ & 
+\frac{\nu^8 \beta^8-6 \nu^6 \beta^6+120 \nu^4 \beta^4
  -5040 \nu^2 \beta^2+362880}{39916800 \beta^{11}}
\nu \beta \cos (\nu \beta)\, .
\end{align}
These expressions are useful to evaluate the integrals
$\mathcal{K}(\beta,\mu,\delta)$, $\mathcal{L}(\beta,\mu,\delta,\rho)$
and $\mathcal{M}(\beta,\mu,\delta,\rho)$ numerically. In the regime
where $\beta\ll 1$, they can be expanded, making use of the Taylor
expansion of the cosine integral function
\begin{align}
  \mathrm{Ci}(\vert \nu \vert \beta)
  =\gamma_{\mathrm{E}}+\ln\left(\vert\nu\vert\beta\right)
  -\frac{\nu^2\beta^2}{4}+\mathcal{O}(\beta^4)\, , 
\end{align}
where $\gamma_{\mathrm{E}}$ is the Euler constant. This gives rise to
\begin{align}
C_3(\nu,\beta) =& \frac{1}{2 \beta^2}
+\frac{\nu^2}{4} \left[2 \log (\vert\nu\vert \beta)
  +2 \gamma_{\mathrm{E}} -3\right]-\frac{\beta^2\nu^4}{48}
+\mathcal{O}(\beta^4),\\
C_5(\nu,\beta) =& \frac{1}{4 \beta^4}
-\frac{\nu^2}{4 \beta^2}+\frac{\nu^4}{288}
\left(-12 \log \beta-12 \gamma_{\mathrm{E}} +25\right)
+\frac{\beta^2 \nu^6}{1440}
+\mathcal{O}(\beta^4),\\
C_7(\nu,\beta) =& 
\frac{1}{6 \beta^6}
-\frac{\nu^2}{8 \beta^4}
+\frac{\nu^4}{48  \beta^2}
+\nu^6\frac{20 \log (\vert\nu\vert \beta)
  +20 \gamma_{\mathrm{E}} -49}{14400}
-\frac{\nu^8 \beta^2}{80640}
+\mathcal{O}(\beta^4),\\
C_9(\nu,\beta) =&
\frac{1}{8 \beta^8}
-\frac{\nu^2}{12 \beta^6}
+\frac{\nu^4}{96 \beta^4}
-\frac{\nu^6}{1440 \beta^2} 
\nonumber\\ &
+\nu^8\frac{-280 \log (\vert\nu\vert \beta)-280
  \gamma_{\mathrm{E}} +761}{11289600}
+\frac{\nu^{10} \beta^2}{7257600}
+\mathcal{O}(\beta^4),\\
C_{11}(\nu,\beta) =& 
\frac{1}{10 \beta^{10}}
-\frac{\nu^2}{16 \beta^8}
+\frac{\nu^4}{144 \beta^6}
-\frac{\nu^6}{2880 \beta^4}
+\frac{\nu^8}{80640  \beta^2}
\nonumber\\ &
+\nu^{10}\frac{2520 \log (\vert\nu\vert \beta)
  +2520 \gamma_{\mathrm{E}} -7381}{9144576000}
-\frac{\nu^{12} \beta^2}{958003200}
+\mathcal{O}(\beta^4),\\
S_4(\nu,\beta) =& 
\frac{\nu}{2 \beta^2}
+\frac{\nu^3}{36} \left[6 \log (\vert\nu\vert \beta)
  +6 \gamma_{\mathrm{E}} -11\right]
-\frac{\nu^5 \beta^2}{240}
+\mathcal{O}(\beta^4),\\
S_6(\nu,\beta) =& 
\frac{\nu}{4 \beta^4}
-\frac{\nu^3}{12 \beta^2}
+\nu^5\frac{-60 \log (\vert\nu\vert \beta)
  -60 \gamma_{\mathrm{E}} +137}{7200}
+\frac{\nu^7 \beta^2}{10080}
+\mathcal{O}(\beta^4),\\
S_8(\nu,\beta) =& 
\frac{\nu}{6  \beta^6}
-\frac{\nu^3}{24 \beta^4}
+\frac{\nu^5}{240 \beta^2}
+\nu^7\frac{140 \log (\vert\nu\vert \beta)
  +140 \gamma_{\mathrm{E}} -363}{705600}
-\frac{\nu^9 \beta^2}{725760}
+\mathcal{O}(\beta^4),\\
S_{10}(\nu,\beta) =& 
\frac{\nu}{8  \beta^8}
-\frac{\nu^3}{36 \beta^6}
+\frac{\nu^5}{480 \beta^4}
-\frac{\nu^7}{10080 \beta^2}
\nonumber\\ &
+\nu^9\frac{-2520 \log (\vert\nu\vert \beta)
  -2520 \gamma_{\mathrm{E}} +7129}{914457600}
+\frac{\nu^{11} \beta^2}{79833600}
+\mathcal{O}(\beta^4),\\
S_{12}(\nu,\beta) =& 
\frac{\nu}{10  \beta^{10}}
-\frac{\nu^3}{48 \beta^8}
+\frac{\nu^5}{720 \beta^6}
-\frac{\nu^7}{20160 \beta^4}
+\frac{\nu^9}{725760 \beta^2}
\nonumber\\ &
+\nu^{11}\frac{27720 \log (\vert\nu \vert \beta)
  +27720 \gamma_{\mathrm{E}} -83711}{1106493696000}
-\frac{\nu^{13}\beta^2}{12454041600}
+\mathcal{O}(\beta^4)\, .
\end{align}
The reason why the integrals have been expanded to such a high order
in $\beta$ is that all negative powers of $\beta$ cancel out in the
integrals $\mathcal{K}(\beta,\mu,\delta)$,
$\mathcal{L}(\beta,\mu,\delta,\rho)$ and
$\mathcal{M}(\beta,\mu,\delta,\rho)$ when $\mu\geq -1$, which only
feature mild (\ie logarithmic) divergence in $\beta$. Plugging the
above formulas into \Eq{eq:calK:CS}, one indeed obtains
\begin{align}
\mathcal{K}\left(\beta,-3,\delta\right) \simeq & 
\frac{1}{2\beta^2}+\frac{1}{200} \left[40 (\delta +1) \log (2 \beta )
  -73 \delta +40 \gamma_\mathrm{E}  (\delta +1)-93\right]
\nonumber \\  &
-\frac{3}{350} \beta ^2 (2 \delta +1)
+\frac{\beta^4}{4725}\left(1+3\delta\right)+\order{\delta^2,\beta^6},\\
\mathcal{K}(\beta,-1,\delta) =& - \ln(2\beta)-\gamma_\mathrm{E}+\frac{7}{4}
-\frac{\delta }{2}-\frac{7 \delta ^2}{48}
+\frac{\beta^2}{10}\left(1+\delta+\frac{5}{6}\delta^2\right)
+\order{\beta^4, \delta^3},\\
\mathcal{K}(\beta,1,\delta) =&\frac{9}{4}\left(1-\delta\right)
-\frac{\beta^2}{2}+\order{\delta^2\ln\delta,\beta^4},\\
\mathcal{K}(\beta,3,\delta) =&\frac{9}{4}\left(1-2\ln \frac{\delta}{2}\right)
+9\delta\ln \delta 
-\left[\frac{9}{4}+\ln(512)\right]\delta{-\frac{\beta^4}{4}}
+\order{\delta^2\ln\delta, \beta^6}\, ,
\end{align}
where the result is also expanded in $\delta$ (explicit expressions
where $\delta$ is left free can be obtained but they are rather
cumbersome). For the integrals $\mathcal{L}(\beta,\mu,\delta,\rho)$
and $\mathcal{M}(\beta,\mu,\delta,\rho)$, one further expands in
$\vert\rho\vert$ (after expanding in $\beta$), so two regimes need to
be distinguished.

The first regime is when $\vert \rho \vert \ll 1$. In this case, the
integrals of interest are
\begin{align}
 \mathcal{L}(\beta,-3,\delta,\rho) =&
 \frac{1}{2\beta^2}+\frac{\ln(2\beta)+\gamma_\mathrm{E}}{5}
 \left(1+\delta\right)-\frac{93+73\delta}{200}
 -\frac{3}{350}\left(1+2\delta\right)\beta^2
\nonumber \\ &
+\left[\frac{\ln(2\beta)+\gamma_\mathrm{E}}{6}-\frac{7}{24}
  +\frac{\delta}{12}\right]\rho^2
-\frac{\beta^2 \rho^2}{60}\left(1+\delta\right)
+\order{\beta^4,\rho^4,\delta^2},\\
  \mathcal{L}(\beta,-1,\delta,\rho) =&
  \frac{7}{4}- \ln (2\beta )- \gamma_\mathrm{E} -\frac{\delta}{2}
  +\frac{\beta^2}{10}(1+\delta)
  +\left(  \frac{\beta ^2}{12}-\frac{3}{8}
  +\frac{3}{8}\delta\right)\rho^2
  \nonumber \\ &
  -\frac{3}{160}  \left[\delta +(2-4 \delta )
    \log \frac{\delta }{2}-1\right]\rho^4
  +\order{\beta^4,Z^6, \beta^4 Z^2, \beta^2 Z^4,\delta^2},\\
 \mathcal{M}\left(\beta,-3,\delta, \rho\right) = &
 \frac{1}{2\beta^2}+\frac{1}{200} \left[40 (\delta +1)
   \log (2 \beta )-73 \delta +40 \gamma_\mathrm{E}  (\delta +1)-93\right]
\nonumber\\ &
+\frac{4\ln(2\beta)+4\gamma_\mathrm{E}-7+2\delta}{8} \rho^2
-\frac{3}{350} \beta ^2 (2 \delta +1)
+\frac{3}{32}(1-\delta)\rho^4
\nonumber \\ & 
-\frac{1+\delta}{20}\beta^2 \rho ^2
+\frac{\beta ^4 (3 \delta +1)}{4725}
+\frac{1}{320} \left[\delta +(2-4 \delta )
  \log \frac{\delta }{2}-1\right]\rho^6
\nonumber \\ &
+\order{\beta^6,\beta^2 \rho^4,\rho^8,\delta^2},\\
\mathcal{M}\left(\beta,-1,\delta, \rho \right) = &
\frac{7}{4}-\ln (2\beta )-\gamma_\mathrm{E} -\frac{\delta}{2}
+\frac{\beta^2}{10}\left(1+\delta\right)
+\left(\frac{\beta ^2}{4}-\frac{9}{8}+\frac{9 \delta }{8}\right)\rho^2
\nonumber \\ &
+\frac{3}{32} \left[1-\delta
  -2 (1-2 \delta ) \log \frac{\delta }{2}\right]\rho^4
+\order{\beta^4,\rho^6,\delta^2}\, .
\end{align}
The fact that we have first expanded in $\beta$ and then in $\rho$
means that those expressions only apply when $\rho\ll 1/\beta$, and
one can check that this is always verified for the cases of interest
in the main text, since $\alpha\beta\ll 1$ and $H R_{\mathrm{obs}}\gg
1$ (namely the observable region contains many Hubble patches at the
time of recombination).

The second regime of interest is defined by the condition $\vert \rho
\vert \gg 1$. In this case, the integrals of interest are
\begin{align}
\mathcal{L}(\beta,-3,\delta,\rho) =& \frac{1}{2\beta^2}
+\left[\frac{\ln( \beta \vert \rho \vert)
    +\gamma_\mathrm{E}}{6}-\frac{11}{36}\right]\rho^2
 +\frac{\ln(\beta \vert \rho \vert)+\gamma_\mathrm{E}-1}{5}\left(1+\delta\right)
 \nonumber \\ &
+\frac{3}{175}\frac{1+2\delta}{\rho^2}
-\frac{\rho^4 \beta^2}{240}-\frac{\rho^2\beta^2}{60}
\left(1+\delta\right)-\frac{3}{350}\beta^2\left(1+2\delta\right)
  \nonumber \\ &
  +\order{\rho^6\beta^4,\rho^{-4},\delta^2},\\
  \mathcal{L}(\beta,-1,\delta,\rho) = & 1-\gamma_{\mathrm{E}}
  -\log(\rho\beta)-\frac{1+\delta}{5\rho^2}
\nonumber  \\ &
+\frac{\rho^2\beta^2}{12}
  -\frac{6}{175 \rho^4}\left(1+2\delta\right)
  +\order{\rho^{-6},\rho^4\beta^4,\beta^2,\delta^2},\\
  \mathcal{L}(\beta,1,\delta,\rho) = &\frac{1}{\rho^2}
  +\frac{4 \left(\delta ^4+4 \delta ^3+7 \delta ^2+6 \delta +3\right)}
  {15  \left(\delta ^2+2 \delta +2\right)\rho ^4}
   {-\frac{\beta^2}{2}}
   +\frac{72(1+2\delta)}{175\rho^6}
   +\order{\rho^{-8},\rho^2\beta^4,\delta^2},\\
   \mathcal{L}(\beta,3,\delta,\rho) =&-\frac{2}{\rho ^4}
   -\frac{16 \left(\delta ^4+4 \delta ^3+7 \delta ^2
     +6 \delta +3\right)}{5  \left(\delta ^2+2 \delta +2\right)\rho ^6}
    {-\frac{\beta^4}{4}+\frac{\rho^2\beta^6}{36}}
    \nonumber \\ &
    +\order{\rho^{-8}}+\order{\beta^6,\delta^2},\\
 \mathcal{M}\left(\beta,-3,\delta, \rho\right) = &
 \frac{1}{2\beta^2} + \frac{\rho^2}{4}\left[-3+2\gamma_\mathrm{E}
   +2\ln(\beta \vert \rho \vert)\right]
+\frac{\gamma_\mathrm{E}+\ln(\beta \vert \rho \vert)}{5}(1+\delta)
-\frac{3}{175}\frac{1+2\delta}{\rho^2}
\nonumber \\ & 
-\frac{\beta^2 \rho^4}{48}
-\frac{1+\delta}{20}\beta^2 \rho^2
-\frac{3}{350}\left(1+2\delta\right)\beta^2
+\order{\beta^4 \rho^6,\rho^{-4},\delta^2},\\
 \mathcal{M}\left(\beta,-1,\delta, \rho\right)= &
 -\ln\left(\beta\vert \rho \vert\right) -\gamma_\mathrm{E}
 +\frac{1+\delta}{5 \rho^2}
 +\frac{\rho^2\beta^2}{4}+(1+\delta)\frac{\beta^2}{10}
 +\order{\rho^{-4},\beta^4,\delta^2}\, .
\end{align}

\bibliographystyle{JHEP}
\bibliography{Discord_Real_Space}

\end{document}